\documentclass[preprint]{aastex61}

\newcommand{\partialr}{\frac{\partial}{\partial r}}

\usepackage{epstopdf}
\usepackage{amsmath}
\usepackage{hyperref}
\usepackage{natbib}
\usepackage{bm}

\received{May 24, 2017}
\revised{May 24, 2017}
\accepted{June 5, 2017}
\submitjournal{ApJ}

\shorttitle{MHD Molecular Tornado Models}
\shortauthors{Kelvin Au and Jason D. Fiege}
\begin{document}

\title{Magnetohydrodynamic Models of Molecular Tornadoes}

\correspondingauthor{Jason D. Fiege}
\email{fiege@physics.umanitoba.ca}

\author{Kelvin Au}
\affil{Department of Physics and Astronomy	\\
University of Manitoba	\\
Winnipeg, Manitoba R3T2N2, Canada}

\author{Jason D. Fiege}
\affil{Department of Physics and Astronomy	\\
University of Manitoba	\\
Winnipeg, Manitoba R3T2N2, Canada}


%
%
\begin{abstract}

Recent observations near the Galactic Centre have found several molecular filaments displaying striking helically-wound morphology, which are collectively known as ``molecular tornadoes.'' We investigate the equilibrium structure of these molecular tornadoes by formulating a magnetohydrodynamic model of a rotating, helically magnetized filament. A special analytical solution is derived where centrifugal forces balance exactly with toroidal magnetic stress. From the physics of torsional Alfv\'{e}n waves, we derive a constraint that links the toroidal flux-to-mass ratio and the pitch angle of the helical field to the rotation laws, which we find to be an important component in describing molecular tornado structure. The models are compared to the Ostriker solution for isothermal, non-magnetic, non-rotating filaments. We find that neither the analytic model nor the Alfv\'{e}n wave model suffer from unphysical density inversions noted by other authors. A Monte Carlo exploration of our parameter space is constrained by observational measurements of the Pigtail Molecular Cloud (Pigtail), Double Helix Nebula (DHN), and Galactic Centre molecular Tornado (GCT). Observable properties such as the velocity dispersion, filament radius, linear mass, and surface pressure can be used to derive three dimensionless constraints for our dimensionless models of these three objects. A virial analysis of these constrained models is studied for these three molecular tornadoes. We find that self-gravity is relatively unimportant, whereas magnetic fields, and external pressure play a dominant role in the confinement and equilibrium radial structure of these objects.

\end{abstract}

\keywords{ISM: clouds --- magnetohydrodynamics (MHD) --- methods: numerical}



\section{Introduction} \label{sec:Introduction}

The Pigtail Molecular Cloud (Pigtail) is an intriguing molecular filament that resides close to the Galactic midplane within the central molecular zone (CMZ). It exhibits a helical structure that was first noticed by \cite{Matsumura2012} from CO $J = 1-0$ data observed by the Nobeyama Radio Observatory 45m radio telescope. The Pigtail helix has a $\sim\,15$\,pc pitch that traces at least two rounds of a helix in $^{12}$CO, extending to an angular size of $\sim\,0.15^{\circ} \times 0.15^{\circ}$ and corresponding to a spatial size of $\sim\,20 \times 20$\,pc$^{2}$ with a mass of $(2-6) \times 10^{5}\,M_{\odot}$ \citep{Matsumura2012}.  \par

The proposed mechanism that triggers the helical morphology of the Pigtail is similar to that of two other molecular tornadoes known as the ``Double Helix Nebula'' (DHN) and the ``Galactic Center molecular Tornado'' (GCT) \citep{Matsumura2012}. A vertical magnetic tube that extends into the Galactic halo is twisted into its helical form. The driver of this twisting motion is hypothesized to be due to the shearing of inner ($x_{1}$) and outer ($x_{2}$) cloud orbits. In this scenario, the vertical magnetic tube is caught between two cloud orbits with different velocities, and the velocity shear twists the magnetic tube. With enough twisting, this torsional Alfv\'{e}n wave causes the magnetic tube to undergo a kink ($m=1$) instability  \citep{Matsumura2012, Jackson1975}, which winds it into its observed corkscrew-like morphology. \citet{Matsumura2012} estimate the Pigtail's magnetic field to be on the order of 1\,mG. \citet{Matsumura2012} support this formation scenario due to observations of the SiO/$^{13}$CO ratio that are indicative of the shock between the cloud orbit interaction and of other molecular gas.  \par

Another molecular tornado was found approximately $100$\,pc from the Galactic Centre (GC) in the infrared by the Spitzer Space Telescope by \citet{Morris2006}. They called it the ``Double Helix Nebula'' (DHN) because of its apparently intertwined double helix structure. The DHN is observed to wind at least $1.25$ complete rounds that stretch approximately 25\,pc in length with its long axis parallel the Galaxy's rotation axis. \citet{Morris2006} propose that the DHN is a magnetohydrodynamic (MHD) torsional Alfv\'{e}n wave propagating away from the Galactic disk, driven by the circumnuclear disk (CND). \citet{Morris2006} estimated the magnetic field strength to be $B = 0.1 - 1$\,mG, depending on the assumed proton density in the region. Unlike the Pigtail Molecular Cloud, the presence of two intertwining strands of the DHN suggest that the twisted field due to the torsional wave has triggered an $m = 2$ instability \citep{Morris2006,Jackson1975}.  \par

\citet{Sofue2007} first reported on the GCT, which was also observed in CO with the Nobeyama Radio Observatory. The GCT extends $170$\,pc vertically from the Galactic plane and was measured to rotate at a velocity of $\sim\,30$\,km\,s$^{-1}$ with an estimated mass of $M \,\sim\,1.2\times 10^{6}\,M_{\odot}$. \citet{Sofue2007} note that the GCT, or at least part of it, consists of two helical strands like the DHN. Interestingly, the authors suggest that the GCT is not gravitationally bound, but that there must be an external force or pressure to keep the filament bound. They suggest that the most plausible mechanism confining the GCT against centrifugal force is the magnetic tension due to a toroidally wrapped field with $B \,\sim\,0.4$\,mG. The GCT was proposed to arise from the same mechanism as that of the DHN -- that a torsional Alfv\'{e}n wave, driven by the epicyclic rotation of a cloud orbiting the Galactic Centre, twists a magnetic flux tube, which then undergoes an $m=2$ MHD instability \citep{Sofue2007}.
\par

There has not yet been a detailed theoretical/numerical study on the structure of molecular tornadoes. In this paper, we consider the equilibrium structure of uniformly and differentially rotating, isothermal, self-gravitating and non-self-gravitating, pressure truncated cylinders threaded by somewhat general helical magnetic fields, analogous to the stationary filaments studied by \citet{Fiege1999a, Fiege1999b}. In our equilibrium models, we also take into account the propagation of torsional Alfv\'{e}n waves. A simple analytical model is developed, followed by more general analytical models, and the latter are constrained using the limited observational constraints that are available \citep{Matsumura2012, Morris2006, Sofue2007}. Our models  provide the equilibrium structure needed to study the instabilities described by \citet{Matsumura2012}, \citet{Sofue2007}, and \citet{Morris2006} that are believed to trigger the coiling of a twisted magnetic tube, which we will explore in a future paper. An especially interesting question that we will address in future work is that of why some molecular tornadoes appear to exhibit an $m=1$ mode of instability, while others appear to be dominated by the $m=2$ mode. Additionally, the equilibrium models developed here, and in our forthcoming stability calculations, will enable us to predict sub-millimetre polarization maps, to hopefully further constrain our models in the future, when such maps become available.
\par

\citet{Hansen1976} provided a theoretical study of the equilibrium and stability of uniformly rotating, isothermal, and infinitely long gas cylinders of finite radius. They noted an interesting feature, in which the density fluctuates non-monotonically as a function of radius. These fluctuations can be seen in Figure 1 of \citet{Hansen1976}, where they were referred to as ``density inversions.'' Density inversions were also found by \citet{Recchi2014}, who studied the equilibrium of uniformly and differentially rotating, non-magnetic, pressure truncated filaments. \citet{Hansen1976} suggested that density inversions are due to a battle between gravitational, centrifugal, and pressure gradient forces in maintaining the filament in equilibrium, and \citet{Recchi2014} proposed that such density profiles must be truncated by an external pressure before any density inversion occurs, because they would otherwise not be physically representative of observed filaments. We adopt this viewpoint in our helically magnetized models, which we truncate whenever the density is seen to increase with radius.
\par

\citet{Kaur2006} presented their findings on differentially rotating, self-gravitating filaments, and have incorporated the effects of a helical magnetic field. They took into account the isothermal and logatropic equations of state, combined with constant velocity, and constant angular momentum rotation laws. The magnetic field in their model is proportional to the square root of the density of the filament. We also examine isothermal models, but with rotation laws that follow a power law in radius, as well as rotation laws that follow a power law in magnetic flux.  Physics of torsional Alfv\'{e}n waves are taken into account in our models, which has not been incorporated in the models of \citet{Kaur2006}. The assumption of a torsional Alfv\'{e}n wave couples the toroidal magnetic field $B_{\phi}(r)$ to the rotation law $\Omega(r)$ (see Section \ref{sec:FieldAngle}). Thus, the physics of torsional Alfv\'{e}n waves provides a realistic constraint on the model, and results in interesting insights into the physics of molecular tornadoes.
\par

Our model of molecular tornadoes begins with the equations of steady state MHD (Section \ref{sec:Theory}). A special analytical solution is presented in Section \ref{sec:AnalyticSolution}. This closed form solution is not general, but it is interesting and does provide some insight into the physics of molecular tornadoes. Various differential rotation laws are described in Section \ref{sec:DifferentialRotation} including a constraint due to torsional Alfv\'{e}n waves. We conduct a virial analysis introduced in Section \ref{sec:VirialAnalysis}. Our numerical method is presented in Section \ref{sec:SquashingTransformation}, and the method for constraining our models by observed quantities is included in Section \ref{sec:Dimensionless Quantities: Comparing Observations with Numerical Models}. Results are shown in Section \ref{sec:Results}, discussed in Section \ref{sec:Discussion}, and summarized in Section \ref{sec:Summary}.

\section{Theory}
\label{sec:Theory}

We assume ideal MHD, as described by the four ideal MHD equations in steady state -- the continuity equation, Cauchy momentum equation, Faraday's Law, and the isothermal equation of state, respectively given by

\begin{subequations}\label{eq:SteadyStateMHDEquations}
\begin{equation}
\bm{\nabla}\bm{\cdot}(\rho\bm{v}) = 0 \label{eq:SteadyStateMHDEquationsContinuity},\\
\end{equation}
\begin{equation}
\rho(\bm{v}\bm{\cdot}\bm{\nabla})\bm{v} = -\bm{\nabla} P - \rho\bm{\nabla}\Phi + \frac{(\bm{\nabla}\bm{\times}\bm{B}) \bm{\times} \bm{B}}{4\pi}, \label{eq:SteadyStateMHDEquationsCauchy}\\
\end{equation}
\begin{equation}
\bm{\nabla} \bm{\times} (\bm{v} \bm{\times} \bm{B}) = 0, \text{ and} \label{eq:SteadyStateMHDEquationsAmpere}\\
\end{equation}
\begin{equation}
P = \sigma^{2}\rho \label{eq:SteadyStateMHDEquationsIsothermal},\\
\end{equation}
\end{subequations}

\noindent where $\rho$ is the density, $\bm{v}$ is the velocity, $P$ is the pressure, $\bm{B}$ is the magnetic field, $\Phi$ is the gravitational potential, and $\sigma$ is the velocity dispersion. We consider the velocity dispersion to contain both thermal and nonthermal contributions: $\sigma = \sqrt{\sigma_{thermal}^{2} + \sigma_{nonthermal}^{2}}$.

For convenience, we define a set of dimensionless quantities, written here with an overscript tilde, where the physical quantities are scaled according to the following scaling laws:

\begin{subequations}
\label{eq:ScaleQuantities}
\begin{equation}\label{eq:scalerho}
\rho = \rho_{0} \tilde{\rho}, \\
\end{equation}
\begin{equation}\label{eq:scaler}
r = \frac{\sigma}{\sqrt{4 \pi G \rho_{0}}} \tilde{r}, \\
\end{equation}
\begin{equation}\label{eq:scaleP}
P = \sigma^{2} \rho_{0} \tilde{P}, \\
\end{equation}
\begin{equation}\label{eq:scaleB}
B = \sigma \sqrt{\rho_{0}} \tilde{B}, \\
\end{equation}
\begin{equation}\label{eq:scalePhi}
\Phi = \sigma^{2} \tilde{\Phi}, \\
\end{equation}
\begin{equation}\label{eq:scaleg}
g = \sqrt{4 \pi G \rho_{0}} \sigma \tilde{g}, \\
\end{equation}
\begin{equation}\label{eq:scalem}
m = \frac{\sigma^{2}}{4 \pi G}\tilde{m}, \\
\end{equation}
\begin{equation}\label{eq:scalev}
v_{\phi} = \sigma \tilde{v}_{\phi}, \\
\end{equation}
and
\begin{equation}\label{eq:scaleOmega}
\Omega = \Omega_{0}\tilde{\Omega} = \sqrt{4\pi G \rho_{0}}\tilde{\Omega}.
\end{equation}
\end{subequations}

\noindent In these scaling relations,  $G$ is the universal gravitational constant, $g$ is the gravitational field, $m$ is the linear mass, $v_{\phi}$ is the rotational speed, and $\Omega$ is the rotational frequency. The subscript 0 denotes the quantities at the filament's radial core ($r=0$). It should be noted that all physical quantities, except $\sigma$, are considered functions of $r$. \textit{Hereafter, the overscript tilde notation is dropped, except where otherwise noted, since model quantities are presented in dimensionless form for most of the paper.}

By assuming cylindrical symmetry ($r,\phi,z$) and rotational motion only, the velocity and magnetic field take the form
\begin{subequations}\label{eq:MHDAssumptions}
\begin{equation}
\bm{v} = v_{\phi}(r)\hat{\phi},
\end{equation}
and
\begin{equation}
\bm{B} = B_{\phi}(r)\hat{\phi} + B_{z}(r)\hat{z}.
\end{equation}
\end{subequations}

\noindent Using the assumption of equation \ref{eq:MHDAssumptions}, the only non-trivial terms of equation \ref{eq:SteadyStateMHDEquations} are in the $\hat{r}$ direction. Both the continuity equation and Faraday's law are satisfied automatically under the assumption of equation \ref{eq:MHDAssumptions}, while the Cauchy momentum Equation simplifies to
\begin{equation}\label{eq:11June2015}
0 = -\frac{dP}{dr} - \rho\frac{d\Phi}{dr} + \rho\frac{v_{\phi}^{2}}{r} - \frac{B_{\phi}^{2}}{4\pi r} - \partialr\bigg(\frac{B_{\phi}^{2}+B_{z}^{2}}{8\pi}\bigg).
\end{equation}

\noindent We introduce the magnetic flux-to-mass ratios formulated by \citet{Fiege1999a}. The toroidal and poloidal magnetic flux-to-mass ratios are defined by
\begin{subequations} \label{eq:FluxToMass}
\begin{equation}\label{eq:FluxToMassphi}
\Gamma_{\phi} = \frac{B_{\phi}}{r\rho},
\end{equation}
and
\begin{equation}\label{eq:FluxToMassz}
\Gamma_{z} = \frac{B_{z}}{\rho},
\end{equation}
\end{subequations}
respectively. The flux-to-mass ratios are free functions of the magnetic flux $\Phi_{M}$ in the context of perfect MHD due to Ferraro's law of isorotation \citep{Ferraro1937}, but flux $\Phi_{M}(r)$ is a function of $r$ in cylindrical symmetry. \citet{Fiege1999a} assumed that both $\Gamma_{z}$ and $\Gamma_{\phi}$ were constant. However, we consider the possibility that these functions might vary as a function of $\Phi_{M}$, or equivalently $r$.

In addition to the equations of MHD, the effects of gravity also need to be taken into account. The gravitational field, $g(r) := -\partial\Phi(r)/\partial r$, is described by Poisson's equation for gravity
,  which becomes 
\begin{equation}\label{eq:PoissonsEquation01}
\frac{dg}{dr} = -\frac{g}{r} - \rho.
\end{equation}
Having defined $g$, equation \ref{eq:11June2015} can be expressed as
\begin{equation}\label{eq:11Oct2015(6)}
\frac{d\rho}{dr} = \frac{\rho g + \rho\Omega^{2}r - \frac{1}{2\pi}r\rho^{2}\Gamma_{\phi}^{2} -\frac{1}{8\pi}r^{2}\rho^{2}\partialr\Gamma_{\phi}^{2}-\frac{1}{8\pi}\rho^{2}\partialr\Gamma_{z}^{2}}{1+\frac{1}{4\pi}r^{2}\rho\Gamma_{\phi}^{2} + \frac{1}{4\pi}\rho\Gamma_{z}^{2}}.
\end{equation}
It is not generally possible to solve this equation analytically except under special assumptions. For most of the work presented in the paper, we apply numerical methods to solve equation \ref{eq:11Oct2015(6)}, taking into account the physics of torsional Alfv\'{e}n waves (see Section \ref{sec:FieldAngle}), which leads to the numerical solutions presented in Section \ref{sec:Results}. However, we first present one special solution that can be solved analytically.

\subsection{Analytic Solution: Rotational Balance with Toroidal Magnetic Stress}
\label{sec:AnalyticSolution}
In this section, we derive a special analytical solution for the case where centrifugal forces exactly balance magnetic stresses from the toroidal field. We acknowledge that this solution is very specific, and therefore lacks the generality of the numerical solutions presented later in this paper. All quantities are presented in dimensionless form, where we have dropped the overscript tilde notation. \par

Various simplifying terms and assumptions are introduced so that the momentum equation (equation \ref{eq:11June2015}) is simplified, which leads to an interesting analytical solution. We first introduce a ``total'' pressure is defined by 

\begin{equation}\label{eq:Ptot1}
P_{tot} = P + \frac{B_{\phi}^{2} + B_{z}^{2}}{8\pi}.
\end{equation}

\noindent Our analytical solution requires that the centrifugal force balances exactly with the inward toroidal magnetic stress, so that

\begin{equation}\label{eq:centrifugalforcebalancetoroidalfield}
\rho v_{\phi}^{2} - \frac{B_{\phi}^{2}}{4\pi} = 0.
\end{equation}

\noindent We further assume that the ratio of pressure and total magnetic pressure is constant: 

\begin{equation}
\label{eq:constantbeta1}
\beta = P\bigg(\frac{B_{\phi}^{2}}{8\pi}+\frac{B_{z}^{2}}{8\pi}\bigg)^{-1} = constant.
\end{equation}

\noindent We define 
\begin{subequations}
\label{eq:constantbetacomponents}
\begin{equation}
\label{eq:constantbetacomponentsz}
\beta_{z}:=\frac{P}{P_{mag,z}},
\end{equation}
and
\begin{equation}
\label{eq:constantbetacomponentsphi}
\beta_{\phi}:=\frac{P}{P_{mag,\phi}},
\end{equation}
\end{subequations}
where $\beta_{z}$ and $\beta_{z}$ are also constant, so that equation \ref{eq:constantbeta1} can be written as 

\begin{equation}
\label{eq:constantbeta2}
\beta = \bigg(\frac{1}{\beta_{z}} + \frac{1}{\beta_{\phi}}\bigg)^{-1}.
\end{equation}

Using the definition of $\beta_{\phi}$ in equation \ref{eq:centrifugalforcebalancetoroidalfield}, and the definitions of $\sigma$ from equation \ref{eq:SteadyStateMHDEquationsIsothermal}, it is straightforward to show that 
\begin{equation}
\label{eq:vphiConstant}
v_{\phi} = \sqrt{\frac{2\sigma^{2}}{\beta_{\phi}}} = constant.
\end{equation}
Equation \ref{eq:vphiConstant} implies that the velocity field is highly sheared at $r=0$, which violates the usual boundary condition  $v_{\phi}(0) = 0$. Thus our special solution may not be physical very close to the axis $r=0$. Interestingly, \citet{Kaur2006} have also considered constant $v_{\phi}$ models. 

The total pressure is then given by the sum of the gas pressure (possibly including non-thermal contributions) and the magnetic partial pressures according to 

\begin{equation}\label{eq:Ptot2}
P_{tot} = P + P_{mag,z} + P_{mag,\phi} = P\bigg(1 + \frac{1}{\beta_{z}} + \frac{1}{\beta_{\phi}}\bigg).
\end{equation}

\noindent Furthermore, we define an \textit{effective} sound speed given by 

\begin{equation}\label{eq:EffectiveSoundSpeed}
\sigma_{eff}^{2} := \bigg(1 + \frac{1}{\beta_{z}} + \frac{1}{\beta_{\phi}}\bigg)\sigma^{2},
\end{equation} 

\noindent where dimensional units are restored for this equation only, 
so that we can \textit{redefine} the dimensionless gravitational potential
\begin{equation}
\label{eq:dimensionlesspotential}
\tilde{\Phi} := \frac{\Phi}{\sigma_{eff}^{2}},
\end{equation}
which is used to express Poisson's equation for gravity in  dimensionless form.  
   \noindent Once again dropping the tilde notation, by assuming that centrifugal force and toroidal magnetic stresses exactly balance (equation \ref{eq:centrifugalforcebalancetoroidalfield}), and incorporating equations \ref{eq:Ptot2} and \ref{eq:dimensionlesspotential}, equation \ref{eq:11June2015} becomes 
\begin{equation}
\label{eq:11June2015Eq5a}
0 = \frac{\partial}{\partial r}\log{\rho} + \frac{\partial}{\partial r}\Phi,
\end{equation}
which can be integrated to find a solution for $\rho$ so that Poisson's equation can be written as  

\begin{equation}
\label{eq:dimensionlesspoisson2}
\nabla^{2} \Phi = e^{-{\Phi}},
\end{equation}

\noindent where $r_{0}$ is redefined as 
\begin{equation}\label{eq:dimensionlessrRedfined}
r_{0} = \frac{\sigma_{eff}}{\sqrt{4\pi G}\rho_{0}}.
\end{equation}

Equation \ref{eq:dimensionlesspoisson2} was solved in cylindrical symmetry by \citet{Ostriker1964a} in his study of non-rotating, non-magnetic, isothermal filaments. Restoring dimensions, we therefore obtain an analogous solution for a rotating, magnetic filament:
\begin{subequations}
\label{eq:11June2015Ostriker}
\begin{equation}
\rho = \frac{\rho_{0}}{\Big(1+\frac{r^{2}}{8r_{0}^{2}}\Big)^{2}},
\end{equation}
and
\begin{equation}
v_{\phi} = \sqrt{\frac{2\sigma^{2}}{\beta_{\phi}}},
\end{equation}
where
\begin{equation}
r_{0}^{2} = \frac{\sigma^{2} (1+\beta_{z}^{-1}+\beta_{\phi}^{-1})}{4\pi G \rho_{0}}.
\end{equation}
\end{subequations}

\noindent Thus, our analytic solution is a rescaling of the Ostriker solution, where the core radius $r_{0}$ is modified by the magnetic field. The filament rotates with constant velocity. The density inversions noted by \citet{Hansen1976} and \citet{Recchi2014} are absent from this special solution. \par
\citet{Nagasawa1987} studied the stability of non-rotating isothermal cylinders threaded by a constant, trivial axial magnetic field. The lack of magnetic stresses from such a constant magnetic field do not modify the equilibrium structure from the Ostriker solution. In contrast, the analytical solution presented above has a radial scale modified by both axial and toroidal magnetic stresses. The solutions we explore throughout the rest of this work include non-trivial magnetic stresses.
\par

\subsection{Differential Rotation}
\label{sec:DifferentialRotation}
The special solution presented in Section \ref{sec:AnalyticSolution} represents a finely tuned special case, which we now generalize. Rotation plays an important role in the morphology of molecular tornadoes as seen in observations \citep{Matsumura2012, Sofue2007, Morris2006} and past work on rotating filaments \citep{Hansen1976, Recchi2014}. It is useful to examine several possible rotation laws in order to accurately describe and understand the physics of molecular tornadoes.

\subsubsection{Angular Frequency as a Power Law}
We consider a rotation law where $\Omega$ scales with the radius $r$ according to a power law with index $\alpha$:

\begin{equation}\label{eq:OmegaPowerLaw}
\Omega = \Omega_{0}\bigg(\frac{r}{r_{0}} \bigg)^{\alpha}.
\end{equation}
\noindent Keplerian rotation and solid body rotation are considered to be limiting cases. We show below that this assumption constrains $\alpha$ to the range $-1 \leq \alpha \leq 0$, where $\alpha = 0$ is the case for solid body rotation. For clarity, we note that the $\alpha=-1$ index would be more accurately described as ``Keplerian-like'' where gas orbits a line mass in the absence of pressure gradients. \par

\subsubsection{Keplerian Rotation}
Poisson's equation and the divergence theorem are used to find the following relationship for $g$ in terms of $r$: $2\pi g r L = 4\pi G M(r)$, where $M$ is the filament mass and $L$ is its length. By solving for $g$ and introducing the linear mass $m = M/L$, equating the angular acceleration and gravitational acceleration implies that 

\begin{equation}
\label{eq:SolidBodyOmega}
\Omega = \frac{v_{\phi}}{r} = \frac{\sqrt{2Gm(r)}}{r}.
\end{equation} 

\noindent Thus we find that Keplerian rotation is just a special case of the radial power law (equation \ref{eq:OmegaPowerLaw}) with $\alpha = -1$.

\subsubsection{Rotational Frequency as a Free Function of Magnetic Field Lines}
\label{sec:PhiB}

We follow the methods introduced by \citet{Mouschovias1976a} and \citet{Tomisaka1988} to describe the rotational frequency as a function the magnetic flux $\Phi_{M}$, so that $v_{\phi} = \Omega(\Phi_{M})r$.
Assuming $\Gamma_{z}$ is constant, we find that $\partial \Phi_{M}/\partial r = 2 \pi \Gamma_{z} \rho r$. 
 The angular frequency $\Omega$ is solved numerically from $\partial \Phi_{M}/\partial r$ along with other quantities mentioned later in Section \ref{sec:SquashingTransformation}. The rotational frequency is then assumed to follow a power law
\begin{equation}\label{eq:OmegaPhiB}
\Omega(\Phi_{M}) = \Omega_{0}\Phi_{M}^{\alpha_{M}}.
\end{equation}
The lower limit on $\alpha_{M}$ is determined by the proportionality relations: $v_{\phi}\propto\Omega r \propto\Phi_{M}^{\alpha_{M}}r$ and $\Phi_{M}\propto B_{z}r^{2}\propto r^{2}$, so that $v_{\phi}\propto r^{2\alpha_{M}+1}$. For $2\alpha_{M}+1>0$, this indicates that $\alpha_{M}>-1/2$. For the sake of interest, we also analyze $\alpha_{M}\geq -1$ to understand how $\alpha_{M}$ affects the behaviour of the radial density. \par

\subsubsection{Torsional Alfv\'{e}n Waves}
\label{sec:FieldAngle}

Appendix \ref{sec:SmallOscillationTorsionalAlfvenWave} derives equations for small amplitude oscillatory torsional Alfv\'{e}n waves. Here, we present an analogous derivation for the wave front of a large amplitude torsional Alfv\'{e}n wave, which is important for the numerical solutions presented in Section \ref{sec:Results}. In both Appendix \ref{sec:SmallOscillationTorsionalAlfvenWave} and this large amplitude scenario, we assume that the unperturbed rotation velocity and magnetic fields are described by
\begin{subequations}\label{eq:FieldAngleAssumptions}
\begin{equation}
\bm{B}_{0} = B_{z,0}\hat{z},
\end{equation}
\begin{equation}
\bm{v}_{0} = 0,
\end{equation}
\begin{equation}
\bm{B}_{1} = B_{\phi,1}(z,t)\hat{\phi},
\end{equation}
and
\begin{equation}
\bm{v}_{1} = v_{\phi,1}(z,t)\hat{\phi}.
\end{equation}
\end{subequations}
\noindent  The MHD equations (equation \ref{eq:SteadyStateMHDEquations}) yield

\begin{equation}\label{eq:WaveEquation}
\partial_{t}^{2}B_{\phi}=v_{A}^{2}\partial_{z}^{2}B_{\phi},
\end{equation}

\noindent which is the general wave equation describing transverse perturbations propagating along the flux tube at the Alfv\'{e}n speed \citep{Jackson1975}

\begin{equation}\label{eq:AlfvenSpeed}
v_{A} = \frac{B_{0}}{\sqrt{4\pi\rho_{0}}}.
\end{equation}

\par Consider an Alfv\'{e}n wave propagating locally along the flux tube, where the rotation velocity and magnetic fields are  finite amplitude perturbations. Furthermore, $B_{z}$ and $\rho$ are constant in $z$ and $t$.
If an initially stationary flux tube starts to rotate at its base at time $t=0$, the perturbation travels a distance $v_{A}dt$ in time $dt$, displacing the flux tube by distance $v_{\phi}dt$. As the wave propagates along the flux tube, a toroidal magnetic field $B_{\phi}$ is generated. Consequently, Faraday's Law  reduces to
\begin{equation}\label{eq:AmperesLaw0}
\frac{dB_{\phi}}{dt} = B_{z}\frac{dv_{\phi}}{dz}.
\end{equation}
 The interface ($z=0$) of the propagating wavefront separates the conditions above ($z>0$) and below ($z<0$) the wavefront. Above the wavefront interface, $B_{z}$ is constant, and $B_{\phi}=0$. Below the interface, the magnetic field has been perturbed and exhibits a toroidal component $B_{\phi}$, and rotates at $v_{\phi}$.  A torsional Alfv\'{e}n wave propagates along a magnetic flux tube at the Alfv\'{e}n speed $v_{A} = dz/dt$. By integrating Faraday's Law (equation \ref{eq:AmperesLaw0}), substituting $dt$ from $v_{A}$, and introducing the aforementioned bounds, we obtain
\begin{equation}
\int_{B_{\phi}}^{0}dB_{\phi} = \frac{B_{z}}{v_{A}}\int_{v_{\phi}}^{0}dv_{\phi},
\end{equation}
\noindent and follow through with the integration. 
Thus the field angle $\theta$ is perturbed according to
\begin{equation}\label{eq:FieldAngle1} 
\tan\theta := \frac{B_{\phi}}{B_{z}} = \frac{v_{\phi}}{v_{A}},
\end{equation}
\noindent  We use the flux-to-mass ratios (equation \ref{eq:FluxToMassphi} and equation \ref{eq:FluxToMassz}) in equation \ref{eq:FieldAngle1}, as well as the Alfv\'{e}n speed ($v_{A}$) to find 
\begin{equation}\label{eq:FieldAngle2}
\Gamma_{\phi} = \sqrt{\frac{4\pi}{\rho}}\Omega,
\end{equation}
\noindent where $\Gamma_{\phi} = \Gamma_{\phi}(r)$, in general. 

\subsection{Virial Analysis} \label{sec:VirialAnalysis}
The work by \citet{Sofue2007} suggests that the GCT may not be bound by self-gravity. By extension, it may be possible that other molecular tornadoes may not require significant self-gravity to hold together. It is helpful to conduct a virial analysis in order to investigate the relative energies associated with various physical properties such as self-gravity, external pressures, magnetic fields, turbulence, and rotation. 
\par
Following \citet{Fiege1999a}, the virial equation that describes our models is given by
\begin{equation}\label{eq:VirialEquation}
2\mathcal{K} + \mathcal{M} + \mathcal{W} = 0,
\end{equation}
\noindent where script characters denote quantities per-unit-length; thus $\mathcal{K}$ is the total kinetic energy per unit length, $\mathcal{M}$ is the magnetic energy per unit length, and $\mathcal{W}$ is the gravitational energy per unit length. The total kinetic energy contains the rotational kinetic energy $\mathcal{K}_{rot}$, surface pressure term $\mathcal{K}_{P}$, and internal turbulence $\mathcal{K}_{\sigma}$ so that
\begin{equation}\label{eq:KineticEnergy}
\mathcal{K} = \mathcal{K}_{rot} + \mathcal{K}_{P} + \mathcal{K}_{\sigma}.
\end{equation}
\noindent The rotational kinetic energy is given by $\mathcal{K}_{rot} = (1/2)\mathcal{I}\Omega^{2}$, where $\mathcal{I}$ is the rotational inertia per unit length of a cylindrical filament spinning along its vertical axis. Energy associated with external pressure is given by $\mathcal{K}_{P} = -(3/2)P_{S}\mathcal{V}$ where $\mathcal{V} = \pi r_{S}^{2}$ is the cross-sectional area of the filament. The energy due to internal turbulence is given by $\mathcal{K}_{\sigma} = (3/2)m\sigma^{2}$. Gravitational energy is described by $\mathcal{W} = -m^{2}G$, which is valid for any equation of state, magnetic field, and internal structure \citep{Fiege1999a}. Rearranging equation \ref{eq:VirialEquation} and normalizing by $|\mathcal{W}|$ yields
\begin{equation}\label{eq:scriptMKPn1}
\frac{\mathcal{M}}{|\mathcal{W}|} = -2\bigg(\frac{\mathcal{K}_{rot}}{|\mathcal{W}|}+\frac{\mathcal{K}_{\sigma}}{|\mathcal{W}|}-\frac{|\mathcal{K}_{P}|}{|\mathcal{W}|}\bigg) - 1,
\end{equation}
\noindent which is useful to find the relative magnetic energy components scaled by the gravitational energy. Determining the relative energies of these properties allows us to gain insight into which ones are particularly important for molecular tornadoes. Equation \ref{eq:scriptMKPn1} makes it clear how terms contribute to the sign of $\mathcal{M}/|\mathcal{W}|$, which determines the dominating component of the magnetic field \citep{Fiege1999a}. For $\mathcal{M}/|\mathcal{W}|<0 $, the toroidal component is dominant in pinching the filament, while $\mathcal{M}/|\mathcal{W}|>0 $ implies net support by the poloidal field. The numerical results in Section \ref{sec:Results} are examined with the help of equation \ref{eq:scriptMKPn1}.
\par

\subsection{Guidelines for Stability}\label{sec:Guidelines for stability}
\cite{Shafranov1956} investigated the stability criterion for MHD instabilities. Unlike our equilibrium model for molecular tornadoes, the filaments discussed by \cite{Shafranov1956} do not include rotation, self-gravity, nor are they truncated by external pressure. Nevertheless, \cite{Shafranov1956} provides a guideline for the stability analysis to be performed on our equilibrium model of molecular tornadoes, which will be presented in a future paper. \par
In the case where there is no $B_{z}$ external to the cylinder, the condition for instability is
\begin{equation}\label{eq:Shafranov1958eq14}
\frac{B_{\phi S}}{B_{z}}=\frac{v_{\phi S}}{v_{A}}>\bigg[{\frac{1}{k^{2}}\bigg(1-\frac{m^{2}}{m+kK_{m-1}(k)/K_{m}(k)}\bigg)\bigg(k\frac{I_{m-1}(k)}{I_{m}(k)}\bigg)}\bigg]^{1/2}
\end{equation}
\noindent\citep{Shafranov1956}, where the magnetic field pitch angle relationship (equation \ref{eq:FieldAngle1}) at the cylinder surface is used, and $I_{m}$ and $K_{m}$ are modified Bessel functions of the first and second kind, respectively.
In the case where there is no $B_{z}$ external to the cylinder, the $m=0$ (sausage), $m=1$ (kink), and $m=2$ modes are unstable when
\begin{subequations}\label{eq:Shafranov1958_stability_m}
\begin{equation}\label{eq:Shafranov1958_stability_m=0}
\frac{B_{\phi S}}{B_{z}}=\frac{v_{\phi S}}{v_{A}}>\sqrt{2},
\end{equation}
\begin{equation}\label{eq:Shafranov1958_stability_m=1}
\frac{B_{\phi S}}{B_{z}}=\frac{v_{\phi S}}{v_{A}}>1,
\end{equation}
and
\begin{equation}\label{eq:Shafranov1958_stability_m=2}
\frac{B_{\phi S}}{B_{z}}=\frac{v_{\phi S}}{v_{A}}\gtrsim\frac{1}{0.26}\approx 3.85,
\end{equation}
respectively.
\end{subequations}
Equivalently, the stability conditions above can be expressed in terms of the rotational period $P=2\pi/\Omega_{S}$, and an Alfv\'{e}n crossing time $\tau_{A}:=\lambda/v_{A}$ that describes the time it takes for an Alfv\'{e}n wave to propagate across one wavelength or pitch $\lambda$. The stability conditions in equation \ref{eq:Shafranov1958_stability_m} for $m=0,1,\text{and }2$, respectively, are expressed as
\begin{subequations}\label{eq:Shafranov1958_stability_P}
\begin{equation}
P<\frac{2\pi}{\sqrt{2}}\frac{r_{S}}{\lambda}\tau_{A},
\end{equation}
\begin{equation}
P<2\pi\frac{r_{S}}{\lambda}\tau_{A},
\end{equation}
and
\begin{equation}
P\lesssim 2\pi(0.26)\frac{r_{S}}{\lambda}\tau_{A}.
\end{equation}
\end{subequations}
Equation \ref{eq:Shafranov1958_stability_m} suggests that filaments that are stable against $m=1$ instabilities are also stable against $m=2$ instabilities. In other words, filaments that spin faster may still be stable against $m=2$, but not $m=1$ instabilities. It may also help to understand the stability conditions based on the axis ratio $r_{S}/\lambda$, as expressed in equation \ref{eq:Shafranov1958_stability_P}. For a given $v_{\phi}/v_{A}$, increasing the axis ratio will first stabilize against $m=2$ instabilities before stabilizing against $m=1$ instabilities. It will be interesting to see how these stability conditions described by \cite{Shafranov1956} will compare to our molecular tornado models.
Although \cite{Shafranov1956}'s models are not exactly like those described in this paper, these stability relations are a useful guideline. In particular, equation \ref{eq:Shafranov1958_stability_m} is used to estimate the angular frequency of our equilibrium models in Section \ref{sec:Dimensionless Quantities: Comparing Observations with Numerical Models} as it sets a lower bound for the angular frequency required for the observed instability to occur. For example, the minimum rotation rate required for a filament to undergo a kink instability would be $\Omega_{S}>\sqrt{2}v_{A}/r_{S}$, according to equation \ref{eq:Shafranov1958_stability_m=1}. In other words, a molecular tornado observed with a kink instability must have been spinning at $\Omega_{S}>\sqrt{2}v_{A}/r_{S}$ for the instability to trigger.

\subsection{Squashing Transformation}
\label{sec:SquashingTransformation}

To efficiently calculate a wide range of $r$, and to better study the large $r$ asymptotic behaviour of our equations, we employ a convenient ``squashing'' transformation to more easily numerically integrate out to large $r$. The squashing transformation transforms the dimensionless radius $r$ into a dimensionless quantity $\xi$ defined by  $\xi \equiv \ln(r)$. The radial differential operators therefore transform according to

\begin{subequations}
\begin{equation}\label{eq:dr}
\frac{d}{dr} = \frac{1}{r} \frac{d}{d\xi},
\end{equation}
and
\begin{equation}
\frac{d^2}{dr^2} = \frac{1}{r^{2}}\bigg(\frac{d^2}{d\xi ^2} - \frac{d}{d\xi}\bigg).
\end{equation}
\end{subequations}

\noindent The equations for $g(r)=-\partial{\Phi(r)}/\partial r$, $dg/dr$ (equation \ref{eq:PoissonsEquation01}), and $d\rho/dr$ (equation \ref{eq:11Oct2015(6)}) are trivially transformed to equivalent equations using $\xi$ as the independent variable, which is what we solve numerically.	\par

In the case of equation \ref{eq:FieldAngle2}, where $B_{\phi}$ is due to a torsional Alfv\'{e}n wave, it is useful to further simplify Equation \ref{eq:11Oct2015(6)}. If $\Gamma_{z}$ is constant, then equation \ref{eq:11Oct2015(6)} can be expressed explicitly in terms of the dimensionless quantities $\Omega$, and $d\Omega/d\xi$ by using the flux-to-mass ratios (equation \ref{eq:FluxToMass}) along with the expression above for $\Omega$ (equation \ref{eq:FieldAngle2}). Algebraic simplification results in the expression
\begin{equation}\label{eq:11Oct2015FA}
\frac{d\rho}{d\xi} = \frac{\rho g r - \Omega\rho r^{2}(\Omega + \frac{d\Omega}{d\xi})}{1+\frac{1}{2}\Omega^{2}r^{2}+\frac{\rho}{4\pi}\Gamma_{z}^{2}},
\end{equation}
\noindent where we have used equation \ref{eq:FieldAngle2} to write $\Gamma_{\phi}$ in terms of $\Omega$.

\subsection{Dimensionless Quantities: Comparing Observations with Numerical Models}
\label{sec:Dimensionless Quantities: Comparing Observations with Numerical Models}

Observations of molecular tornadoes measure some useful global quantities that can be used to constrain our solutions to equation \ref{eq:11Oct2015FA}. We denote quantities at the surface of the filament with the subscript $S$, the radius from the filament axis to its surface $r_{S}$, surface pressure $P_{S}$, linear mass $m_{S}$, and rotational frequency $\Omega_{S}$. These observables are written in terms of the dimensionless quantities of our model in the same manner as the scaling laws of equation \ref{eq:ScaleQuantities}. Thus, there are five unknowns in the scaling laws: $\tilde{r}_{S}$, $\tilde{P}_{S}$, $\tilde{m}_{S}$, $\tilde{\Omega}_{S}$, and $\rho_{0}$. The readily observable (or at least somewhat observationally constrained) quantities  are $\sigma$, $r_{S}$, $m_{S}$, and $P_{S}$. The central density $\rho_{0}$ cannot be easily obtained observationally, which also makes $r_{0}$ difficult to obtain directly. Thus, we use the definition of $r_{0}$ to eliminate $\rho_{0}$ from the equations. This reduces the scaling laws for $\tilde{r}_{S}$, $\tilde{P}_{S}$, $\tilde{m}_{S}$, $\tilde{\Omega}_{S}$, and $\rho_{0}$ to three equations that describe combinations of the dimensionless quantities on the left side, in terms of observables on the right:

\begin{subequations}\label{eq:NonDimPars}
\begin{equation}\label{eq:r2SPS}
\tilde{r}_{S}^{2}\tilde{P}_{S}= \frac{4 \pi G}{\sigma^{4}} r^{2}_{S}P_{S}, \\
\end{equation}
\begin{equation}\label{eq:mS2}
\tilde{m}_{S} = \frac{4 \pi G}{\sigma^{2}} m_{S}, \\
\end{equation}
and
\begin{equation}\label{eq:Omega2SPS}
\frac{\tilde{\Omega}_{S}^{2}}{\tilde{P}_{S}} =  \frac{\sigma^{2}}{4\pi G} \frac{\Omega_{S}^{2}}{P_{S}}.
\end{equation}
\end{subequations}

The surface pressure can be reasonably constrained from estimates of the isothermal equation of state, as well as from the dynamics and surface density of stars \citep{Rathborne2014,Swinbank2011}. \citet{Rathborne2014} and \citet{Swinbank2011} suggest that $P_{S}/k \gtrsim 10^{7}$\,K\,cm$^{-3}$, and can possibly be as high as $\sim\,10^{8}$\,K\,cm$^{-3}$.  The velocity dispersion for molecular clouds in the CMZ are roughly $\sigma \,\sim\,15$\,km\,s$^{-1}$ \citep{Rathborne2014}. 
The angular frequency of the filament in equilbrium is estimated from the Shafranov stability relations of equation \ref{eq:Shafranov1958_stability_m} as well as from the rotation mechanism, where possible.

\subsubsection{Pigtail, DHN, and GCT}
\label{sec:PigtailDHNGCT}
Our exploration focusses on the Pigtail to constrain our models. Equation \ref{eq:NonDimPars} shows that there are five parameters that are necessary to constrain our dimensionless models by observations, namely  $\tilde{r}_{S}$, $\tilde{P}_{S}$, $\tilde{m}_{S}$, $\tilde{\Omega}_{S}$, and $\sigma$. Most of these parameters were derived by observations of the Pigtail. The only exception was for rotation, which was not detected due to resolution limitations \citep{Matsumura2012}. The Pigtail's velocity dispersion is $\sigma = 13.2$\,km\,s$^{-1}$, and $n(H_{2}) = 10^{3.5\pm 0.25}$\,cm$^{-3}$ ($\sigma$ and $n(H_{2})$ provided via private communications with Dr. Tomoharu Oka, Keio University; see also \cite{Matsumura2012}). Assuming hydrogen abundance $X = 0.7\pm 0.05$, we estimate via the isothermal equation of state that $P_{S}/k = (2\pm 1)\times 10^{8}$\,K\,cm$^{-3}$.  A direct measurement of the rope of the Pigtail from the figures of \citet{Matsumura2012} suggests that its radius is between $2-4$\,pc. The pressure and velocity dispersion are assumed to be constant, and so it may be reasonable that the volume of the Pigtail rope between equilibrium and its present form is conserved. Thus, the radius of the filament in equlibrium $r_{S}$ and the radius of the filament when the Pigtail has undergone instability $r^{\prime}_{S}$, are related by the pitch angle via $r_{S}=r'_{S}\cos(\theta)$. The rotation speed of the equilibrium filament can be estimated either by considering that the flux tube is spun up by the shearing of two cloud orbits \citep{Matsumura2012} or by the stability calculations of \cite{Shafranov1956} mentioned in Section \ref{sec:Guidelines for stability}. In this case, we opt for the shearing mechanism described by \citep{Matsumura2012} to derive the rotation frequency as the Shafranov stability conditions have underlying assumptions that differ from our models. Although, using the $m=1$ stability criterion (equation \ref{eq:Shafranov1958_stability_m=1}) with the equilibrium radius and Alfv\'{e}n speed $v_{A}\sim 35$\,km\,s$^{-1}$ \citep{Matsumura2012} produces a very close estimate (well within one order of magnitude) of the angular frequency to that derived by considering the shearing mechanism. These parameters result in dimensionless constraints 
\begin{subequations}\label{eq:NonDimParsPigtail}
\begin{equation}\label{eq:r2SPSPigtail}
0.61\lesssim\tilde{r}_{S}^{2}\tilde{P}_{S}\lesssim 7.32, \\
\end{equation}
\begin{equation}\label{eq:mS2Pigtail}
1.73\lesssim\tilde{m}_{S}\lesssim 14.63, \\
\end{equation}
and
\begin{equation}\label{eq:Omega2SPSPigtail}
0.96\lesssim\frac{\tilde{\Omega}_{S}^{2}}{\tilde{P}_{S}}\lesssim 11.52.
\end{equation}
\end{subequations}
\par
There is more uncertainty in the physical parameters of the DHN and GCT, however. In particular, $\sigma$, $\Omega_{S}$, and $P_{S}$ are unclear. Generally, molecular clouds of the CMZ exhibit $\sigma\,\sim\,15$ km s$^{-1}$ \citep{Rathborne2014} which is similar to the Pigtail's velocity dispersion. For the DHN and GCT, we consider $10^{7}$ K cm$^{-3}$ $\lesssim P_{S} \lesssim 10^{8}$ K cm$^{-3}$ \citep{Rathborne2014} as a conservative estimate of the surrounding pressure. This range is slightly lower than the value that we have calculated for the Pigtail, and this discrepancy may be due to the density of the Pigtail, which could be lower than $n(H_{2}) = 10^{3.5\pm 0.25}$\,cm$^{-3}$ \citep{Matsumura2012} on its surface, and/or a hydrogen abundance larger than $X = 0.7$.  Our analysis focusses primarily on the Pigtail because observational measurements for the Pigtail have been more conclusive. Still, we attempt to carry out our analysis with the DHN and GCT, but the results should be taken with more caution. \par
CO observations have revealed two molecular counterparts that are likely associated with the DHN at radial velocities of $\sim-35$ km s$^{-1}$ and $0$ km s$^{-1}$ having lengths $\sim\,150$\,pc, and masses of $0.8 \times 10^{4}\,M_{\odot}$ and $3.3 \times 10^{4}\,M_{\odot}$, respectively \citep{Torii2014, Enokiya2014a}. The equilibrium radius of the DHN filament is conservatively estimated to fall between the narrowest strand width and the overall structure width $\sim 3.5$\,pc \citep{Morris2006}. The CND rotation is the proposed mechanism which drives the DHN. We assume that the CND rotation lies in the range $70 - 110$\,km\,s$^{-1}$ \citep{Morris2006, Enokiya2014a}. Alternatively, with $r_{S}$ and $v_{A}=10^{3}$\,km\,s$^{-1}$ \citep{Morris2006}, an estimate of the angular frequency can be made via the $m=2$ Shafranov stability relation of equation \ref{eq:Shafranov1958_stability_m=2}. We opt to use the angular frequency determined from the CND rotation for the analysis however, since there are assumptions made in equation \ref{eq:Shafranov1958_stability_m=2} that differ from our models (see Section \ref{sec:Guidelines for stability}). Regardless, the general qualitative conclusions drawn from either method of determining the rotation are similar (see Section \ref{sec:IsSelfGravityImportant}). These parameters along with the pressure estimate from \citet{Swinbank2011} and \citet{Rathborne2014}, correspond to 
\begin{subequations}\label{eq:NonDimParsDHN}
\begin{equation}\label{eq:r2SPSDHN}
2\times 10^{-4}\lesssim\tilde{r}_{S}^{2}\tilde{P}_{S}\lesssim 0.27, \\
\end{equation}
\begin{equation}\label{eq:mS2DHN}
0.08\lesssim\tilde{m}_{S}\lesssim 0.32, \\
\end{equation}
and
\begin{equation}\label{eq:Omega2SPSDHN}
80\lesssim\frac{\tilde{\Omega}_{S}^{2}}{\tilde{P}_{S}}\lesssim 3\times 10^{5}.
\end{equation}
\end{subequations}
\par
Assuming $\sigma = 15$ km s$^{-1}$, and provided that the GCT has a density of $\rho \,\sim\,2.7\times 10^{-21}$ g cm$^{-3}$ \citep{Sofue2007}, the isothermal equation of state suggests an internal pressure of $4.4 \times 10^{7}$ K cm$^{-3}$. As a rough guideline, this is within the bounds of the external pressure that we consider. \citet{Sofue2007} draws the similarity between the DHN and GCT that they both exhibit two strands wound in a double helix configuration, and postulates that they arise from a similar mechanism. Thus, we assume that the radius of the GCT is approximately $7-20$\,pc wide, as estimated from the current radius of the GCT, and the size of the molecular cloud it may have originated from \citep{Sofue2007}. The GCT's angular frequency is estimated from the \citet{Shafranov1956} instability condition (equation \ref{eq:Shafranov1958_stability_m=2}). We acknowledge that the Shafranov conditions of equation \ref{eq:Shafranov1958_stability_m} describe minimum bounds for instability, and that the rotation does not have an upper bound. Realistically however, there must be an upper rotation limit otherwise the filament would be unstable and the filament would not remain intact. Even if the GCT were rotating faster than the rate described by equation \ref{eq:Shafranov1958_stability_m=2}, it should not be much faster before undergoing instability. Thus, we conservatively estimate limits on the angular frequency resulting from equation \ref{eq:Shafranov1958_stability_m=2} by multiplying by a factor of 0.5 to 2. These parameters correspond to 

\begin{subequations}\label{eq:NonDimParsGCT}
\begin{equation}\label{eq:r2SPSGCT}
0.11 \lesssim\tilde{r}_{S}^{2}\tilde{P}_{S}\lesssim 8.71, \\
\end{equation}
\begin{equation}\label{eq:mS2GCT}
1.48\lesssim\tilde{m}_{S}\lesssim 12.04, \\
\end{equation}
and
\begin{equation}\label{eq:Omega2SPSGCT}
1.70\lesssim\frac{\tilde{\Omega}_{S}^{2}}{\tilde{P}_{S}}\lesssim 2\times 10^{3}.
\end{equation}
\end{subequations}
\par

Most of the dimensionless constraints of equations \ref{eq:NonDimParsPigtail}, \ref{eq:NonDimParsDHN}, and \ref{eq:NonDimParsGCT} are based on absolute minimum and maximum bounds of the observed measurements, except where otherwise noted. These quantities are chosen to conservatively estimate the bounds of equation \ref{eq:NonDimPars} for each molecular tornado, in an attempt to account for the uncertainty in the observations (particularly with uncertainties associated with the DHN, and GCT).

\section{Results}
\label{sec:Results}
Results of numerically solving equations \ref{eq:11Oct2015(6)} and \ref{eq:11Oct2015FA}, along with $d\Phi/d\xi$ and $dg/d\xi$ , and the rotation laws  described in Section \ref{sec:DifferentialRotation} are presented in this section. The density $\rho$ is calculated from the general form of the Cauchy momentum equation (equation \ref{eq:11Oct2015(6)}) for models excluding the Alfv\'{e}n wave constraint (equation \ref{eq:FieldAngle2}). Equation \ref{eq:11Oct2015FA} was derived  under the constraint of equation \ref{eq:FieldAngle2}. The radial profile of $\rho$ changes due to different rotation laws (equations \ref{eq:OmegaPowerLaw}, and \ref{eq:OmegaPhiB}), magnetic fields, and self-gravity. These profiles are compared to the Ostriker solution, equation \ref{eq:11June2015Ostriker}, for isothermal, non-rotating, non-magnetic, self-gravitating cylinders as a benchmark. At large radii, the Ostriker solution behaves as $\rho \,\sim\,r^{-4}$. 

In our exploration, we are free to explore many different filament models. We have the flexibility to include or exclude self-gravity, choose between various rotation laws and their respective power law indices (see Section \ref{sec:DifferentialRotation}), and define the allowed range of $r$, $\Omega_{0}$, $\Gamma_{z}$, and $\Gamma_{\phi}$. While all of these models are interesting to explore, the most physically relevant models make use of the magnetic field angle constraint from torsional Alfv\'{e}n waves, presented in Section \ref{sec:FieldAngle}. However, we also explore models that exclude this magnetic field constraint for comparison.

Numerical integration was performed with MATLAB's ode45, which is a Runge-Kutta 4-5 integrator. In a few cases where stiffness was detected, we used MATLAB's ode23s, which is a low order method based on the Rosenbrock formula, and efficiently solves rapidly changing problems. We integrate over the range  $r_{min}\leq r \leq r_{max},$ where the lower and upper limits to the dimensionless radius are $r_{min} = 10^{-6}$ and $r_{max} = 10^{6}$, respectively.
If $\Omega$ follows the power law rotation given by equation \ref{eq:OmegaPowerLaw}, then $\Omega_{0}$, $\Gamma_{z}$, and $\Gamma_{\phi}$ are free parameters. In this case, $\Omega_{0}$, $\Gamma_{z}$, and $\Gamma_{\phi}$ are each randomly chosen over the intervals 
\begin{subequations}\label{eq:PLParRange}
\begin{equation}
0 \leq \Omega_{0} \leq 12,
\end{equation}
\begin{equation}
10^{-3} \leq \Gamma_{z} \leq 100,
\end{equation}
\text{and}
\begin{equation}
0 \leq \Gamma_{\phi} \leq 100.
\end{equation}
\end{subequations}
The limits of equation \ref{eq:PLParRange} are conservatively chosen, with the limits of $\Gamma_{\phi}$ and $\Gamma_{z}$ extended beyond that of \cite{Fiege1999a} to account for the environment of the Galactic Centre. Each random set of $\Omega_{0}$, $\Gamma_{z}$, and $\Gamma_{\phi}$ define a particular untruncated filament model. Otherwise if $\Omega$ follows a power law of magnetic flux (equation \ref{eq:OmegaPhiB}), then $\Gamma_{z}$, and $\Gamma_{\phi}$ are free, randomly chosen over the same interval is in equation \ref{eq:PLParRange}, and define a unique density profile. 
The initial conditions at $r_{min}$ are
\begin{subequations}\label{eq:InitialConditions}
\begin{equation}
\rho_{0} = 1,
\end{equation}
\begin{equation}
\Phi_{0} = 0,
\end{equation}
\begin{equation}
g_{0} = 0,
\end{equation}
\begin{equation}
\Phi_{M,0} = \Gamma_{z}\pi r_{min}^{2} \rho_{0},
\end{equation}
and
\begin{equation}
m_{0} = \pi \rho_{0} r_{min}^{2}.
\end{equation}
\end{subequations}
The reader is reminded that all quantities in our numerical calculations are dimensionless.

\begin{figure*}
\figurenum{1}
\gridline{\fig{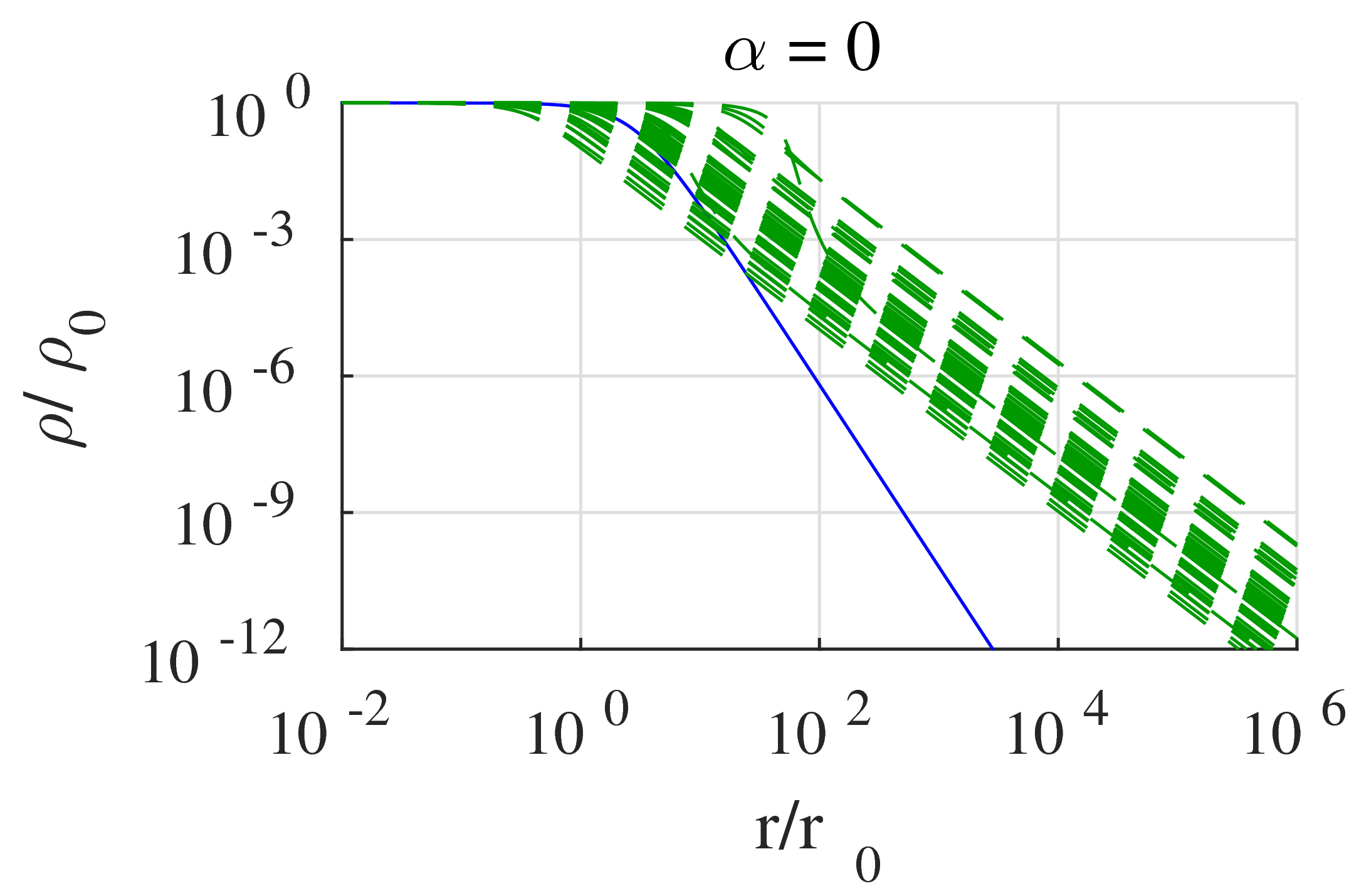}{0.3\textwidth}{}
		  \fig{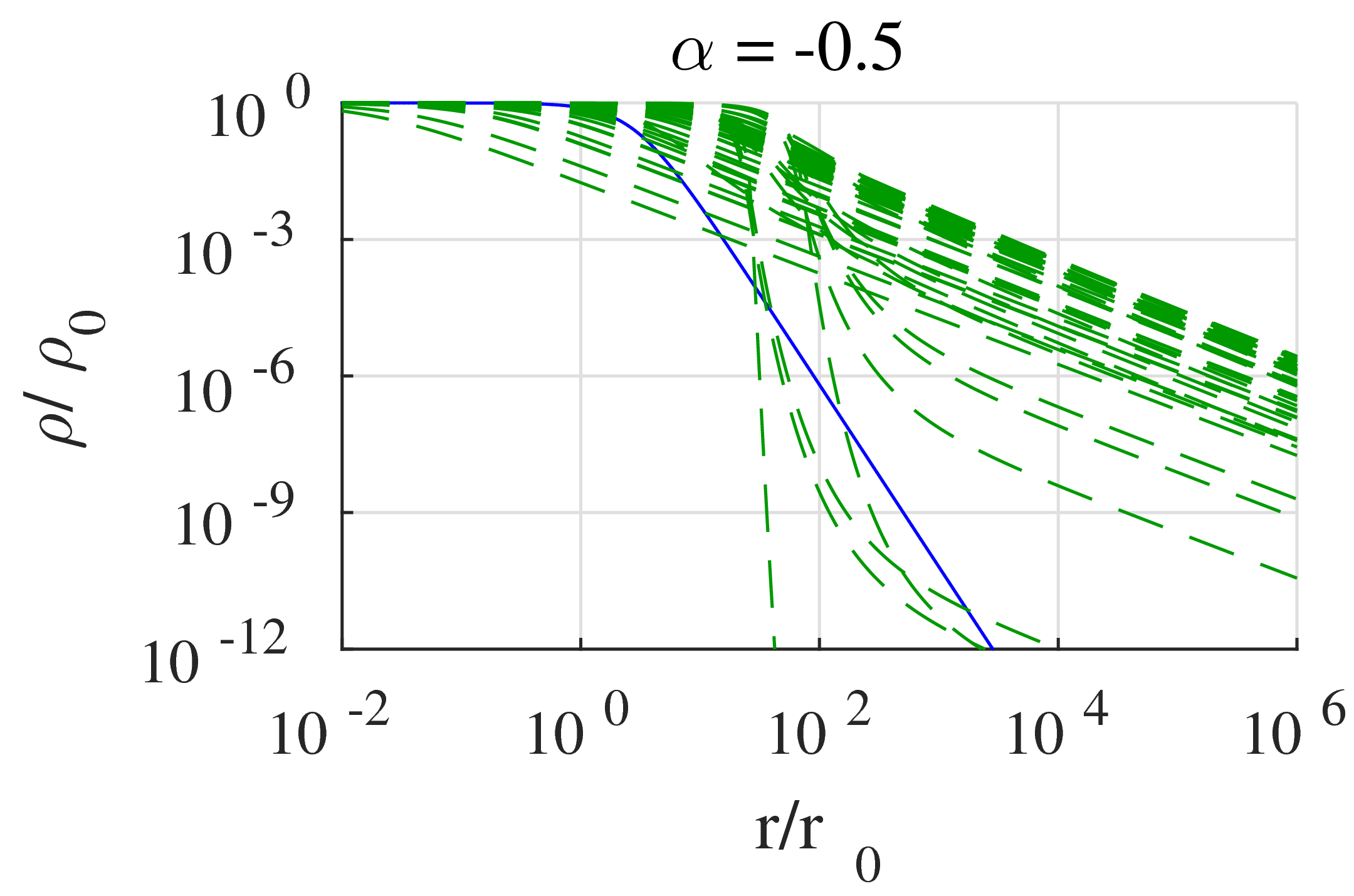}{0.3\textwidth}{}
		  \fig{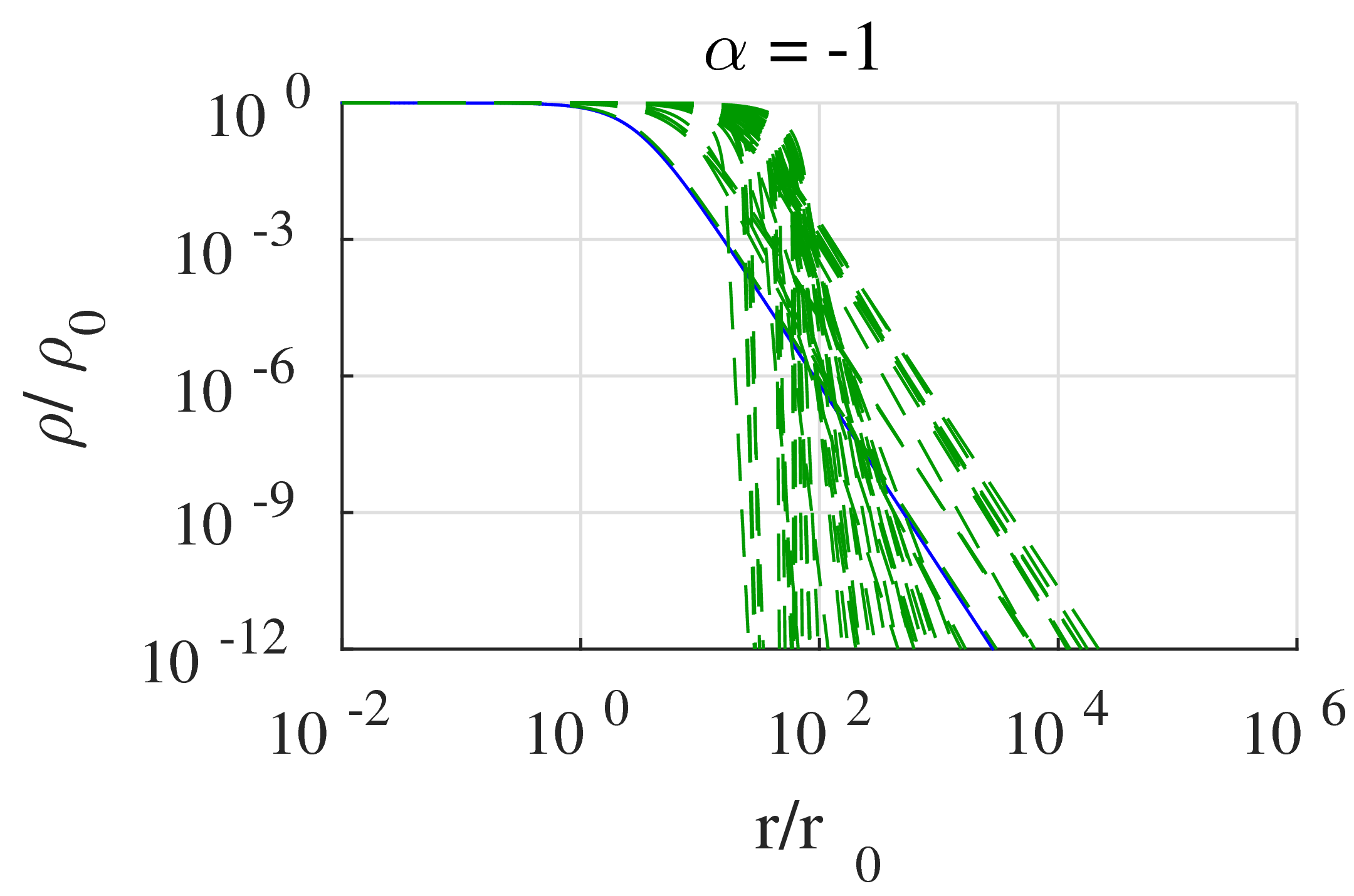}{0.3\textwidth}{}
		  }
\caption{\textbf{Density profiles for filaments obeying the rotation law $\mathbf{\Omega = \Omega_{0}(r/r_{0})^{\alpha}}$ and the torsional Alfv\'{e}n wave constraint}. Green, dashed lines are theoretical density profiles. The Ostriker solution is the solid blue line for all plots, henceforth. Notice that as the power law index becomes more negative, solutions that behave as $\rho \sim r^{-4}$ or steeper occur more frequently.  In each plot, there are 50 density profiles. It should be noted that each plot is magnified for clarity and the integration is from $r_{min}\leq r \leq r_{max}$.  \label{fig:DensityProfilesTheoreticalPL}}
\end{figure*}

\begin{figure*}
\figurenum{2}
\gridline{\fig{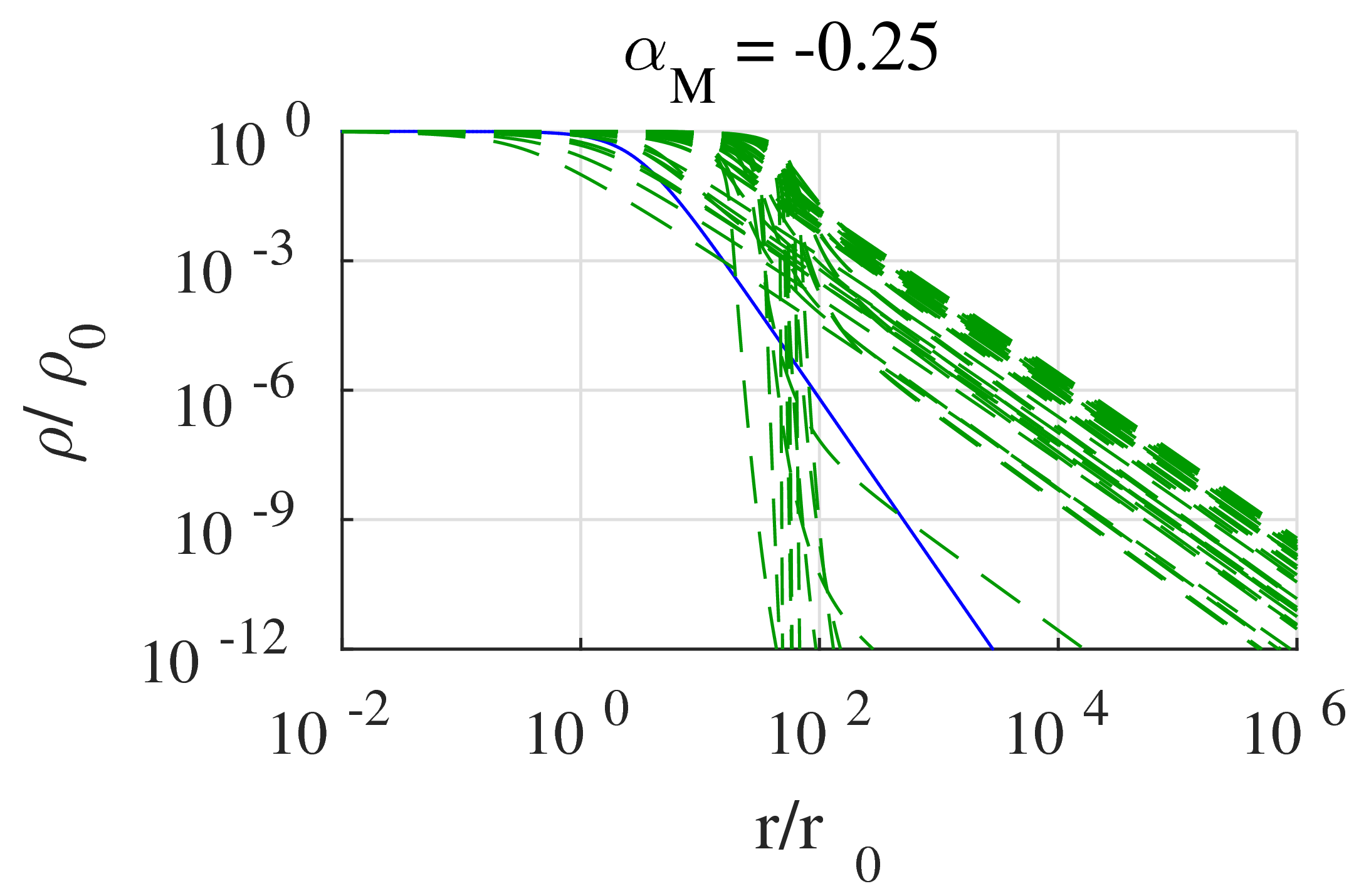}{0.3\textwidth}{}
		  \fig{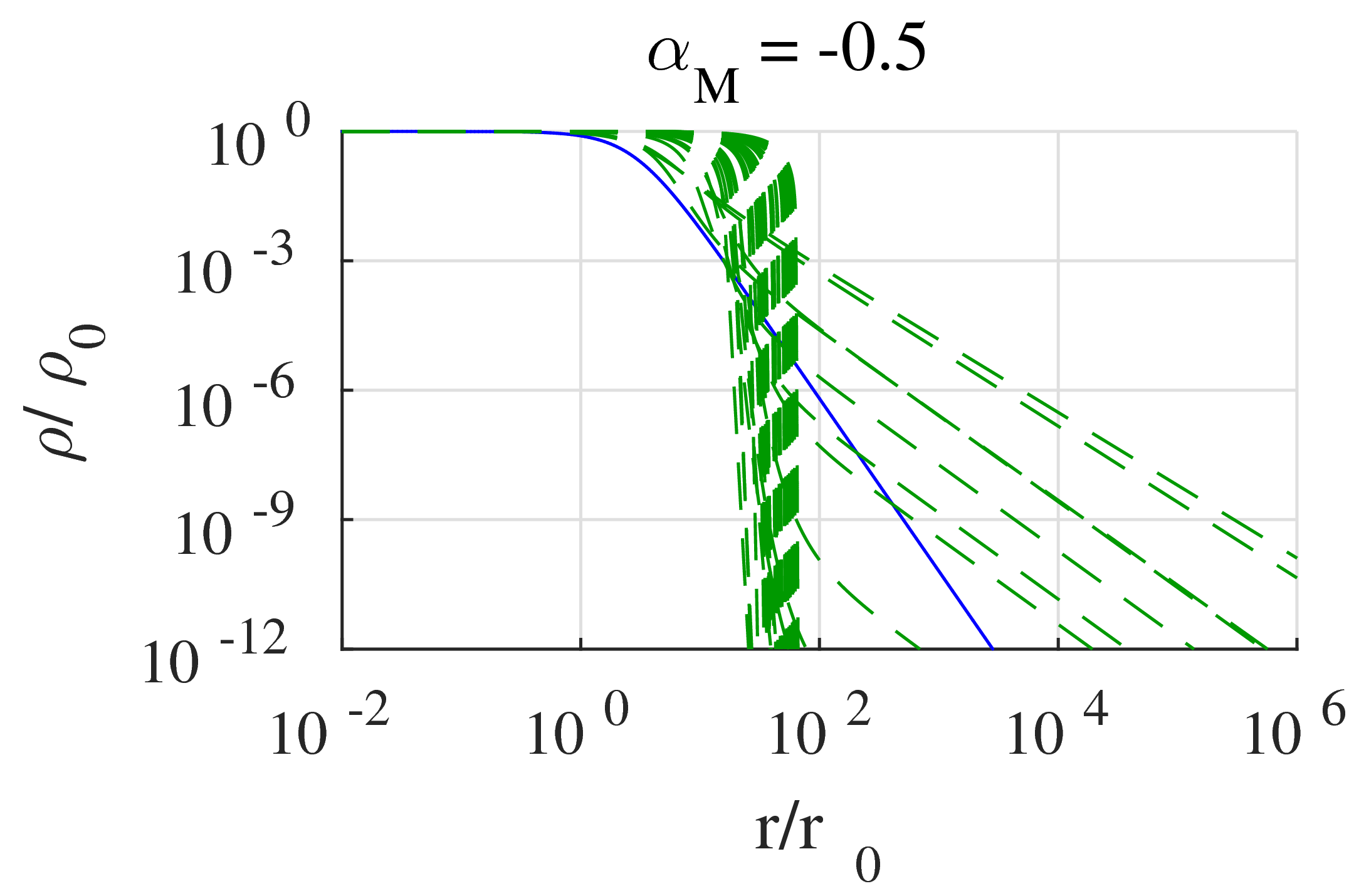}{0.3\textwidth}{}
		  \fig{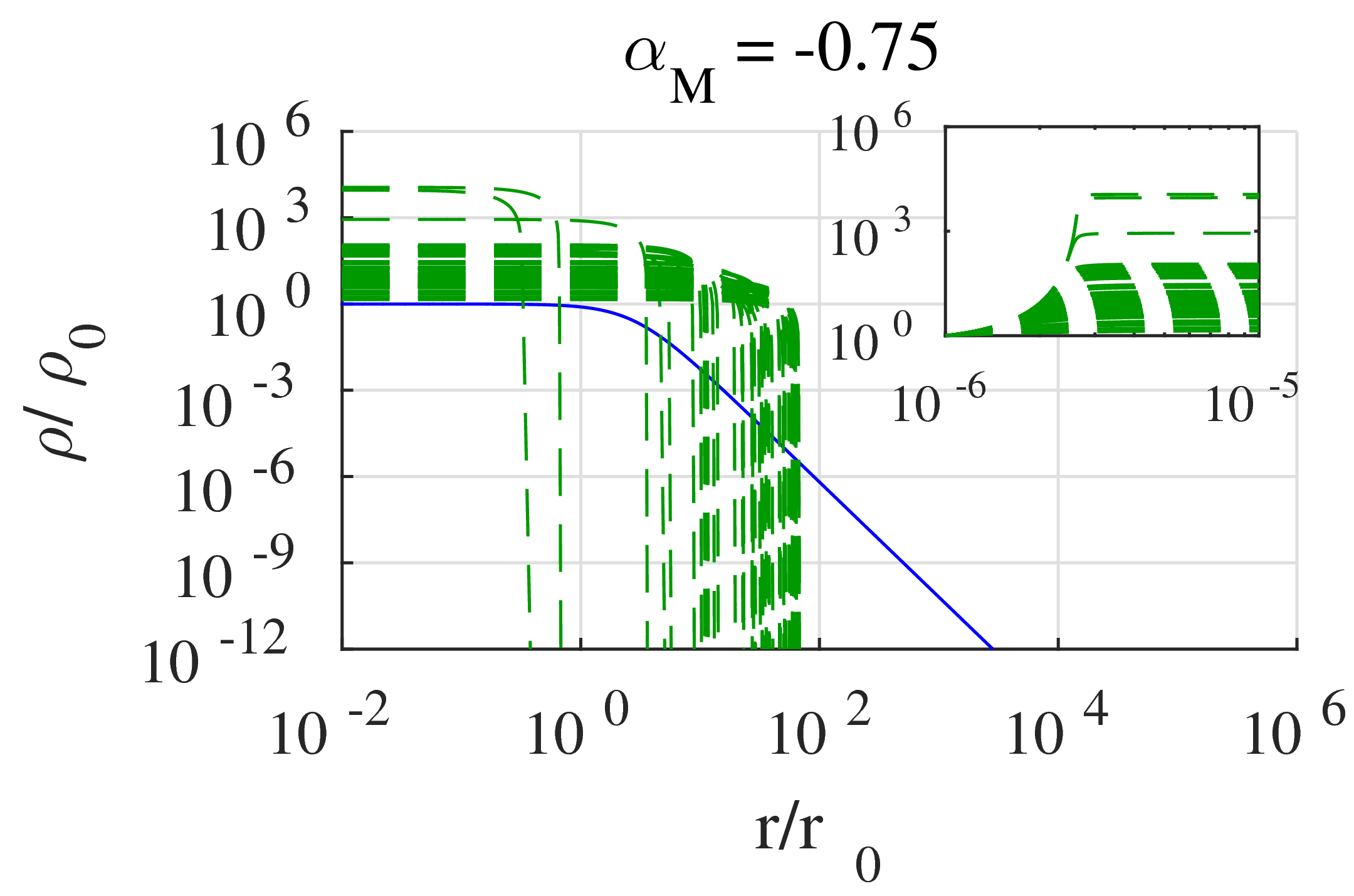}{0.3\textwidth}{}
		 }
\caption{\textbf{Density profiles for filaments obeying the rotation law $\mathbf{\Omega = \Omega_{0}\Phi_{M}^{\alpha_{M}}}$ and the torsional Alfv\'{e}n wave constraint} Green, dashed lines are theoretical density profiles. Notice that as the power law index becomes more negative, solutions can become steeper.  In each plot, there are 50 density profiles. It should be noted that each plot is magnified for clarity and the integration is from $r_{min}\leq r \leq r_{max}$. The inset has the same axis quantities as the host plot, and show the boundary condition $\rho(r_{min})/\rho_{0}=1$. The extreme density inversions shown in the inset are indeed part of the solutions, which we note are unphysical density inversions, and therefore excluded as feasible models. These extreme density inversions are due to $\Omega$ behaving as $r^{-1}$ or steeper, so there is extreme rotation near the origin. The $\alpha_{M}=-0.75$ plot is included in this figure for the sake of interest only.  \label{fig:DensityProfilesTheoreticalPhiB}}
\end{figure*}

\begin{figure*}
\figurenum{3}
\gridline{\includegraphics[width=0.3\textwidth]{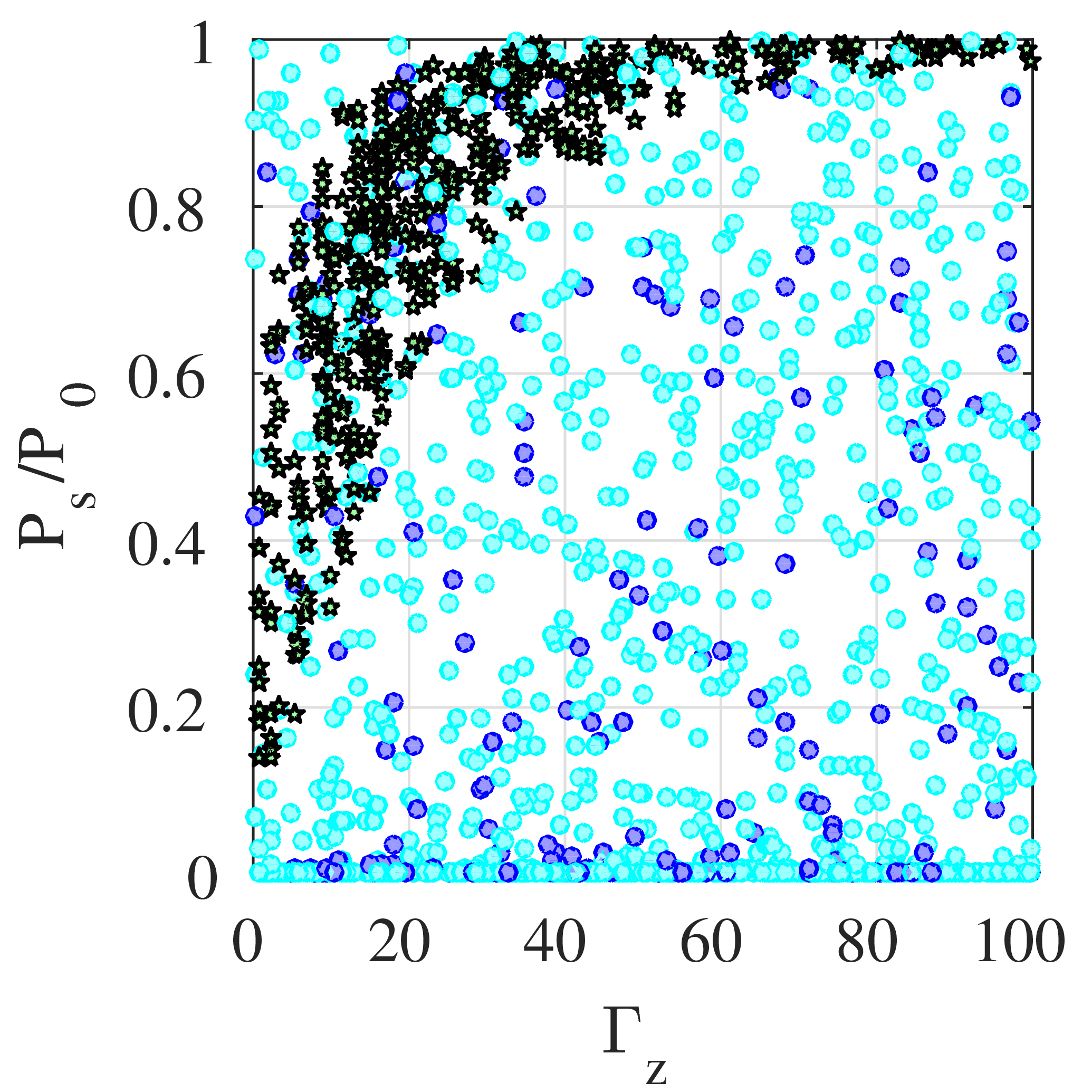}
          \includegraphics[width=0.3\textwidth]{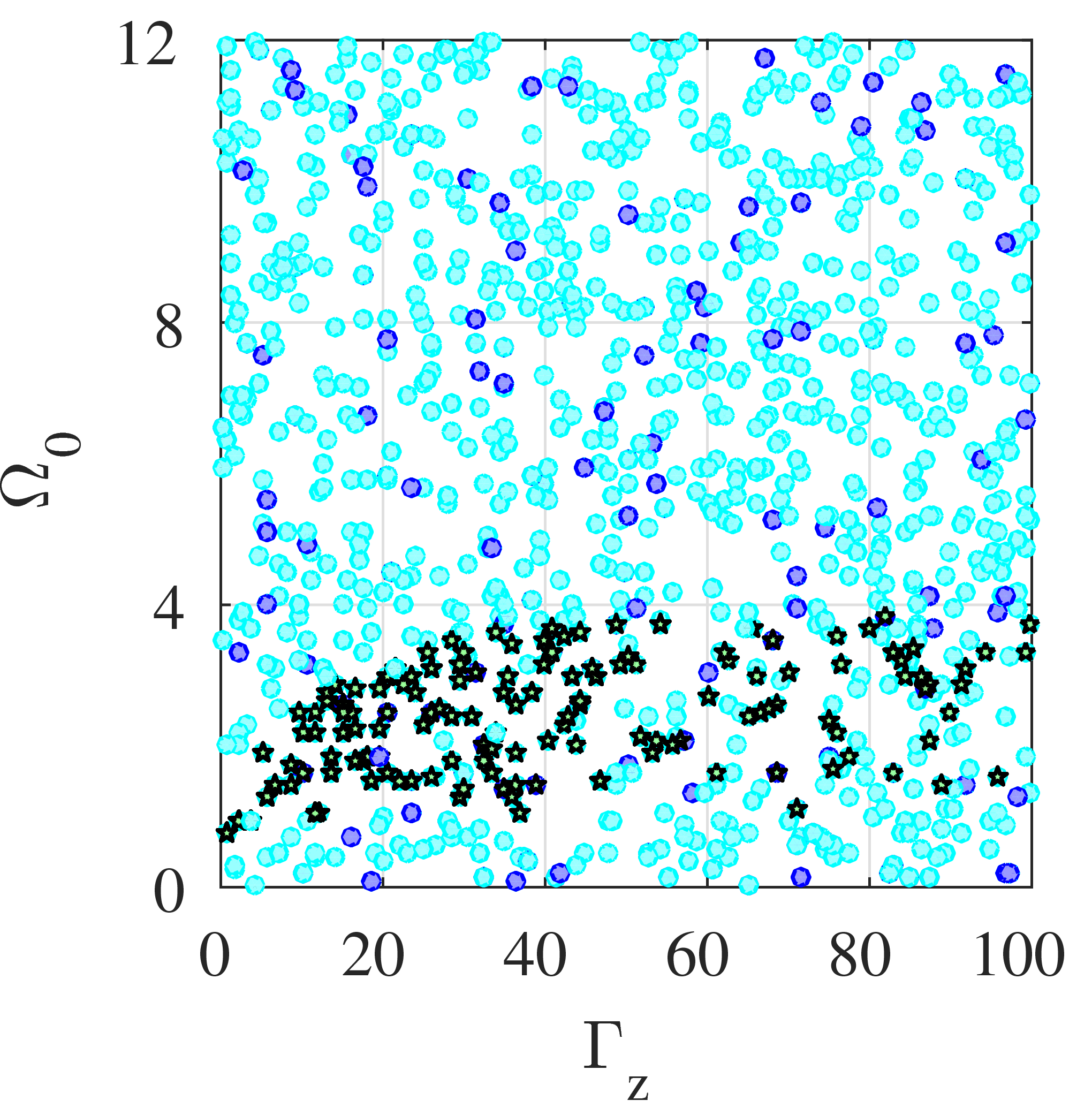}
          \includegraphics[width=0.3\textwidth]{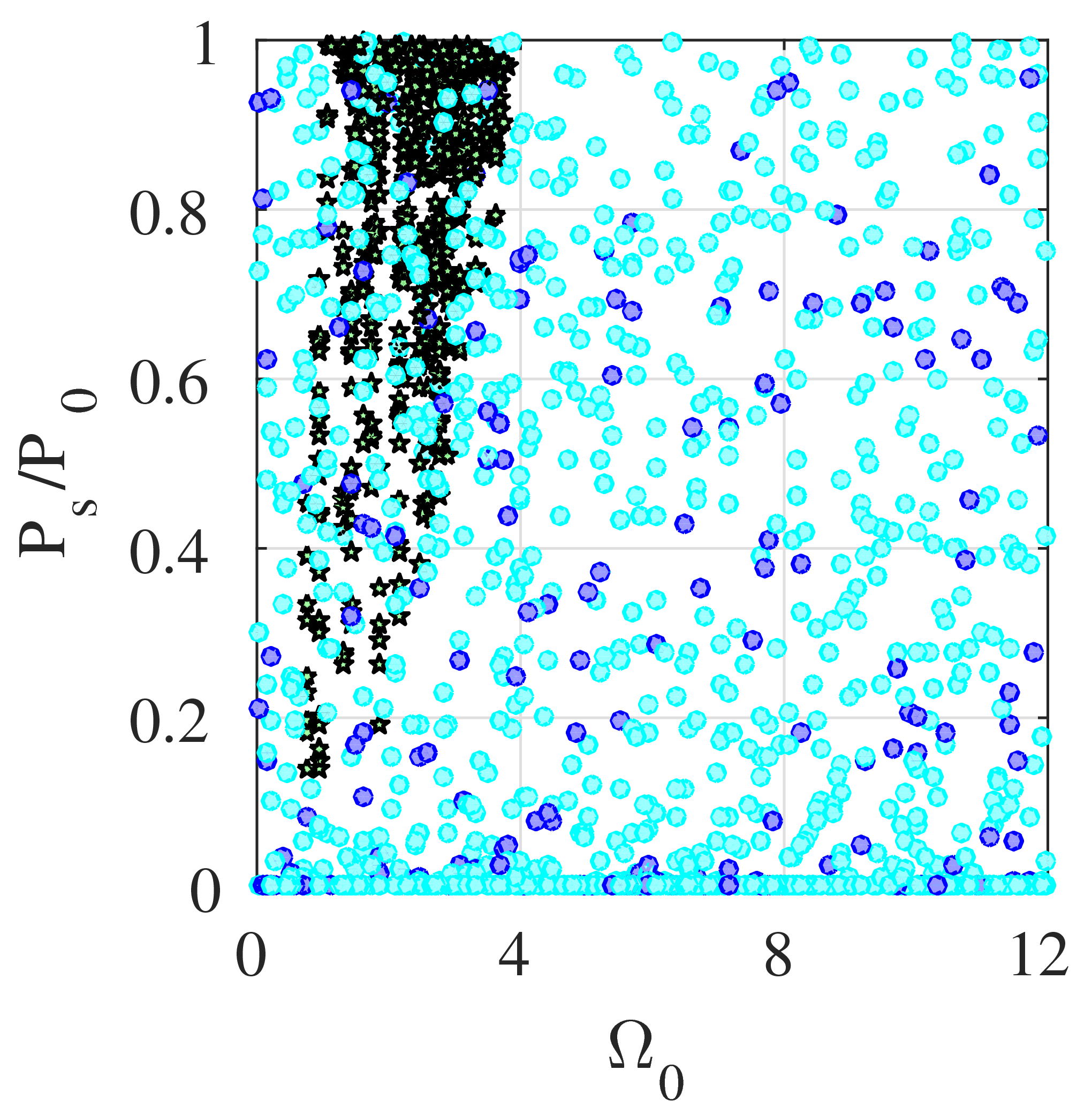}
         }
\caption{\textbf{Observationally constrained solutions}. \textbf{}Scatter plots of the self-gravitating, field-angle constrained, power law rotation (equation \ref{eq:OmegaPowerLaw}) with $\alpha = -0.25$ model parameter space for the Pigtail.  Green stars with black borders indicate where observations reside in the parameter space,  cyan dots indicate models that self-truncate ($d\rho/dr < 0$), and blue dots indicate models where the density reaches asymptotically constant values ($d\rho/dr = 0$).    \label{fig:3DScatterFAPL}}
\end{figure*}

\begin{figure*}
\figurenum{4}
\centering
\plotone{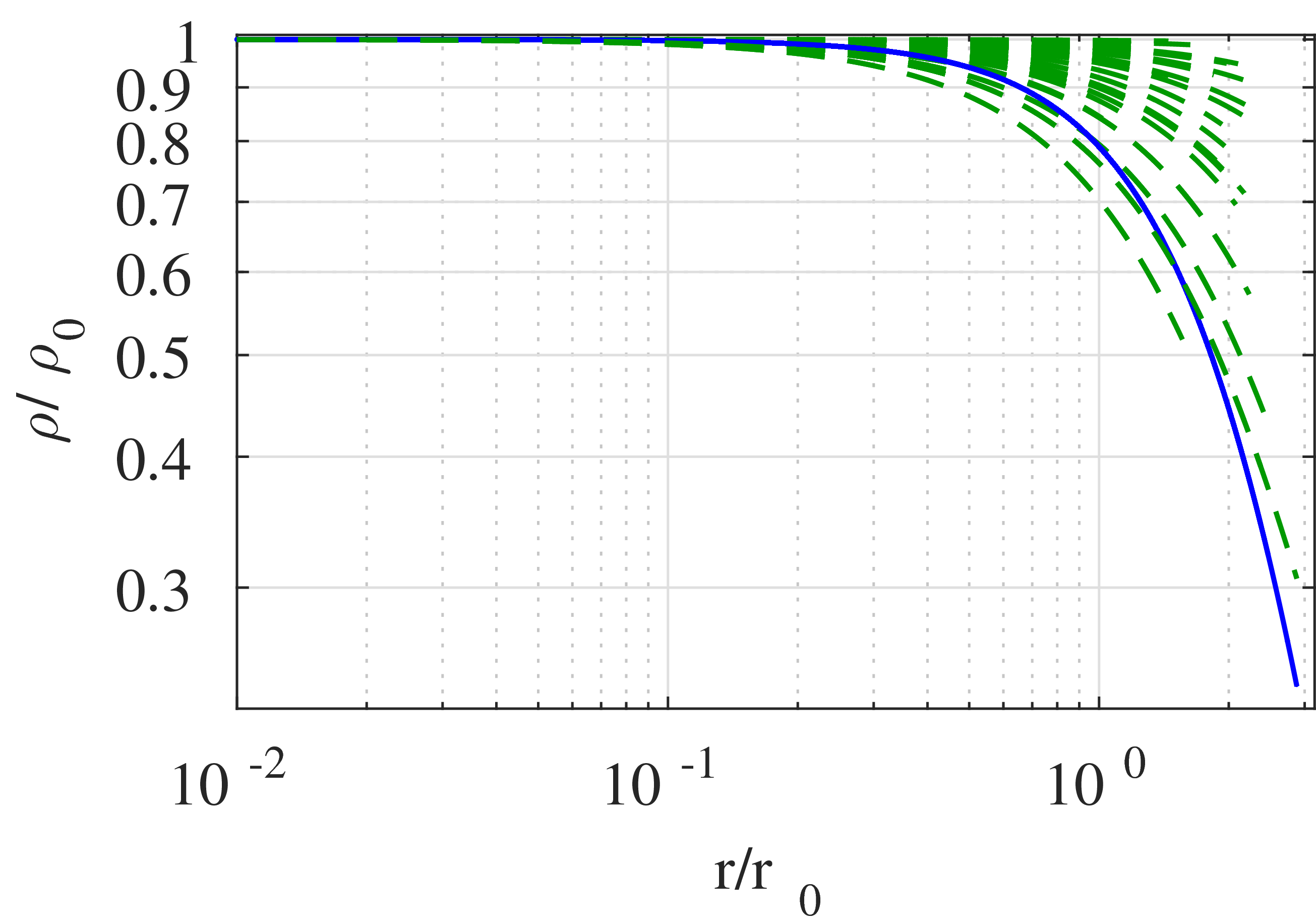}
\caption{\textbf{A sample of density profiles (green dashed lines) corresponding to the green stars in the scatter plots of Figure \ref{fig:3DScatterFAPL}.} Solutions begin at $r/r_{0} = 10^{-6}$ and are pressure truncated around $r/r_{0}=10^{0}$. This figure suggests that external pressure is important in the description molecular tornado structure.  \label{fig:Figure4}}
\end{figure*}

\begin{figure*}
\figurenum{5}
\gridline{\fig{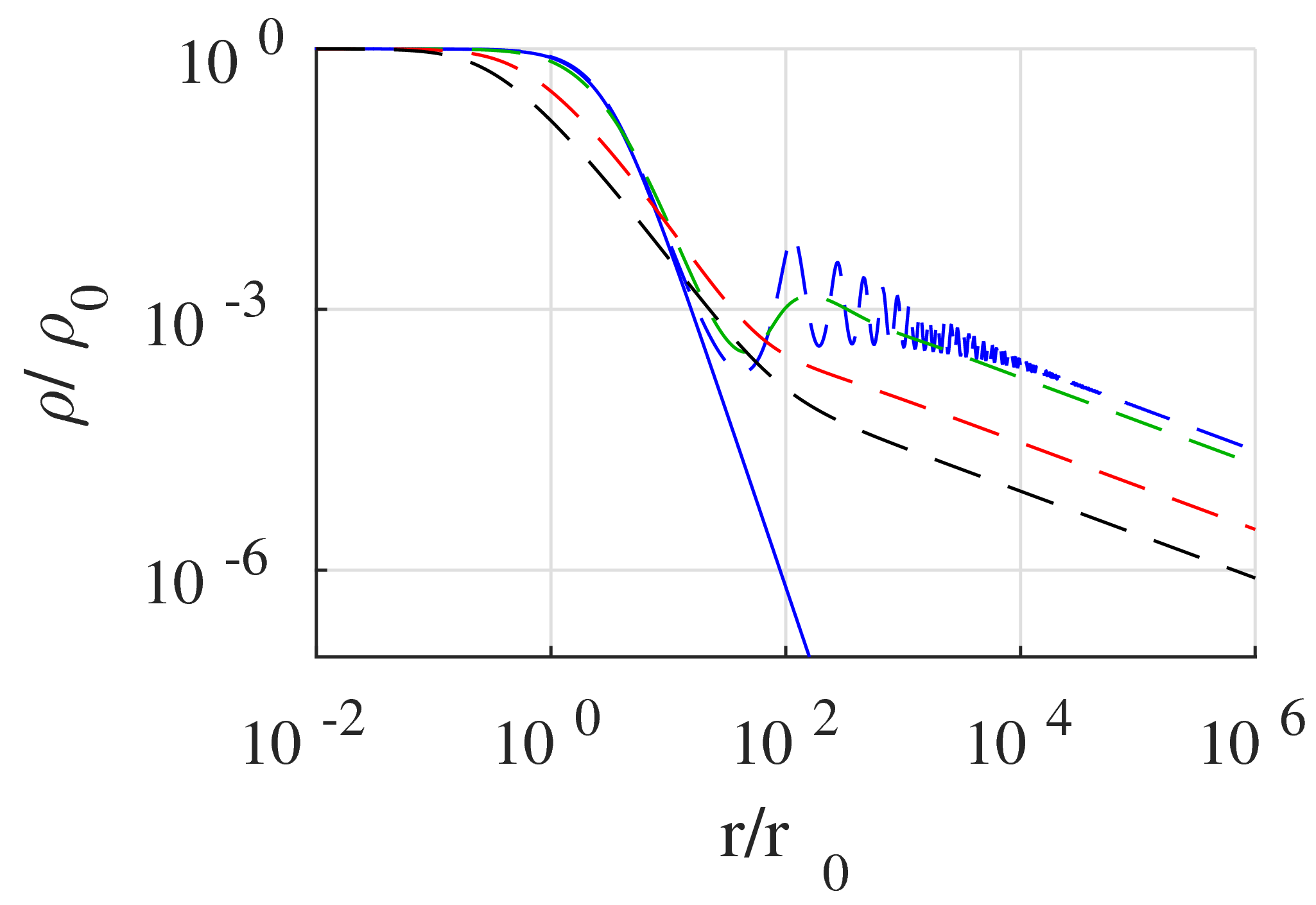}{0.3\textwidth}{}
          \fig{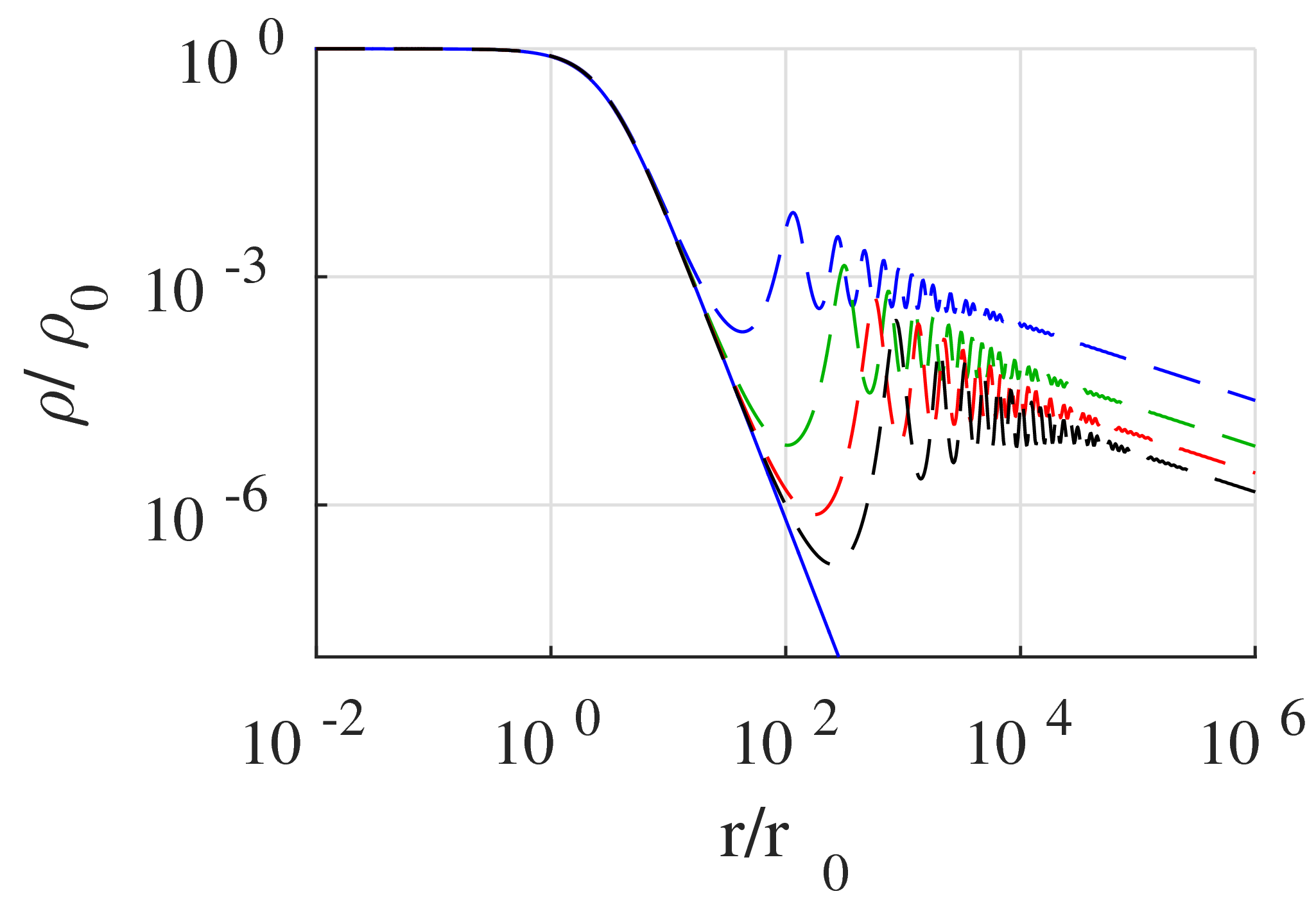}{0.3\textwidth}{}
          \fig{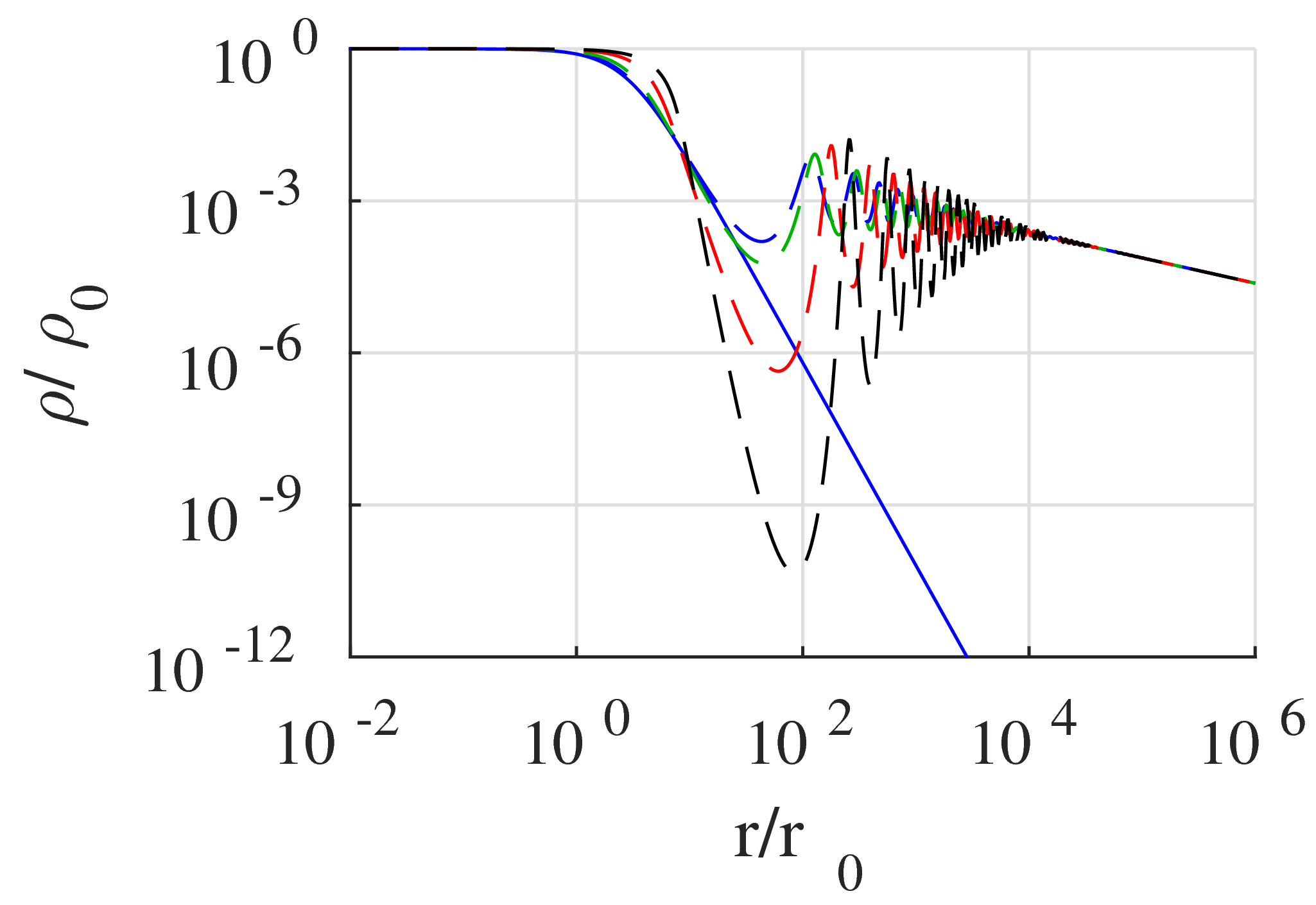}{0.3\textwidth}{}
          }
\caption{\textbf{Density inversions for self-gravitating, radial power law rotating (equation \ref{eq:OmegaPowerLaw}), $\mathbf{\alpha = -0.25}$ model without the torsional Alfv\'{e}n wave constraint of equation \ref{eq:FieldAngle2}}. Constant parameters are set to $\Gamma_{z} = 1$, $\Gamma_{\phi} = \pi/50$, and $\Omega_{0} = \pi/25$. Changing parameters are ordered as the dashed blue, green, red, and black lines, respectively: \textbf{(Left)} $\Gamma_{\phi} = \pi/100, \pi/2, 2\pi, 4\pi$; \textbf{(Centre)} $\Omega_{0} = \pi/25, \pi/50, \pi/75, \pi/100$; \textbf{(Right)} $\Gamma_{z} = \pi/2, \pi, 2\pi, 3\pi$. It should be noted that each plot is magnified for clarity and the integration is indeed from $r_{min}\leq r \leq r_{max}$. Notice that as $\Gamma_{\phi}$ increases, the density profile becomes more pinched, and the inversions smooth out. As $\Omega_{0}$ decreases, the inversions begin at lower densities. A larger $\Gamma_{z}$ causes the inversions to increase in amplitude. Such density inversions are unrealistic, and the density profiles shown here are not truncated for the purpose of illustration only.  \label{fig:161008_fa1sg1_pl_alphan025_dp_inversion}}
\end{figure*}

\begin{figure*}
\figurenum{6}
\gridline{\fig{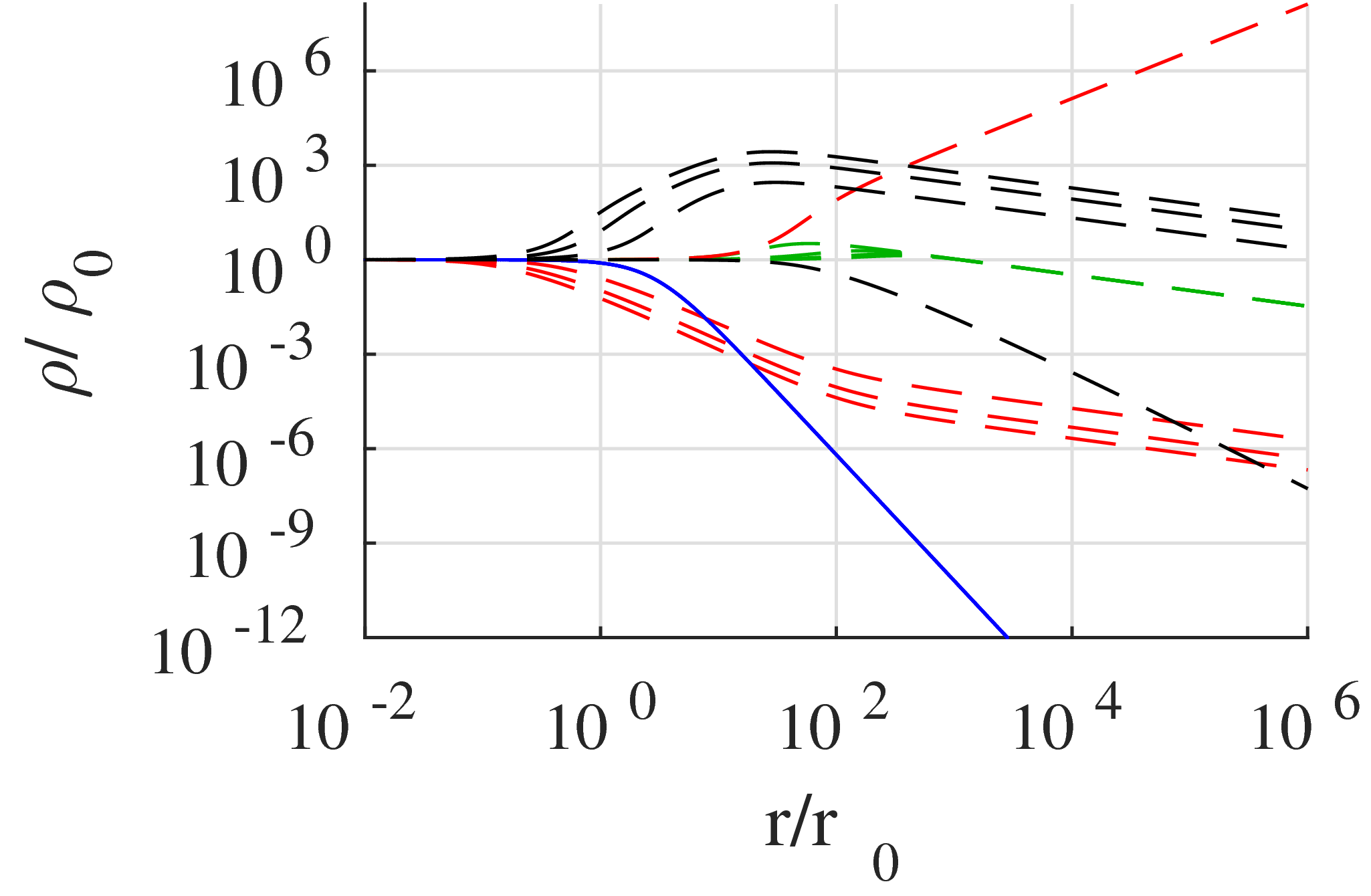}{0.4\textwidth}{}
          \fig{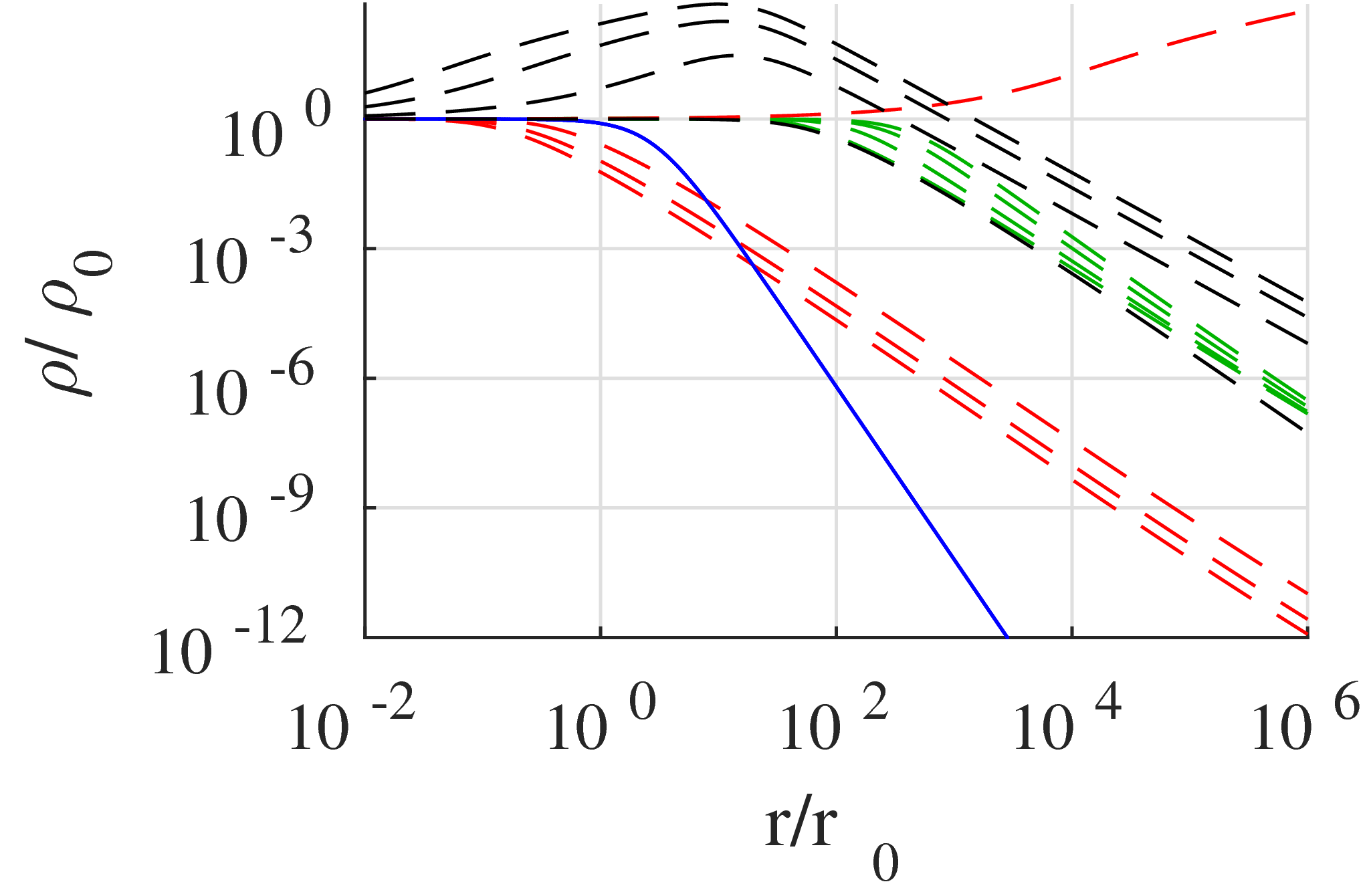}{0.4\textwidth}{}
          }
\gridline{\fig{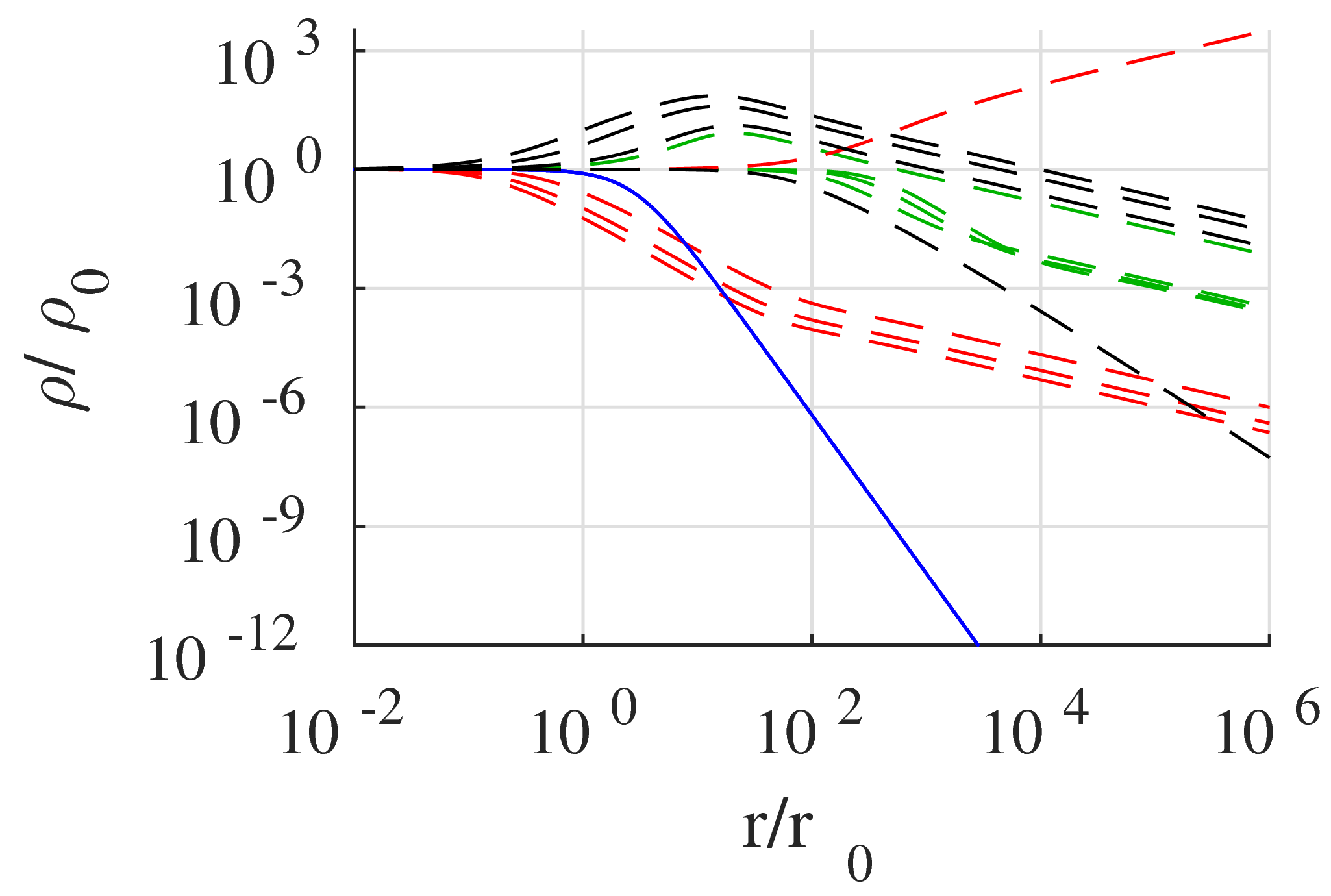}{0.4\textwidth}{}
          \fig{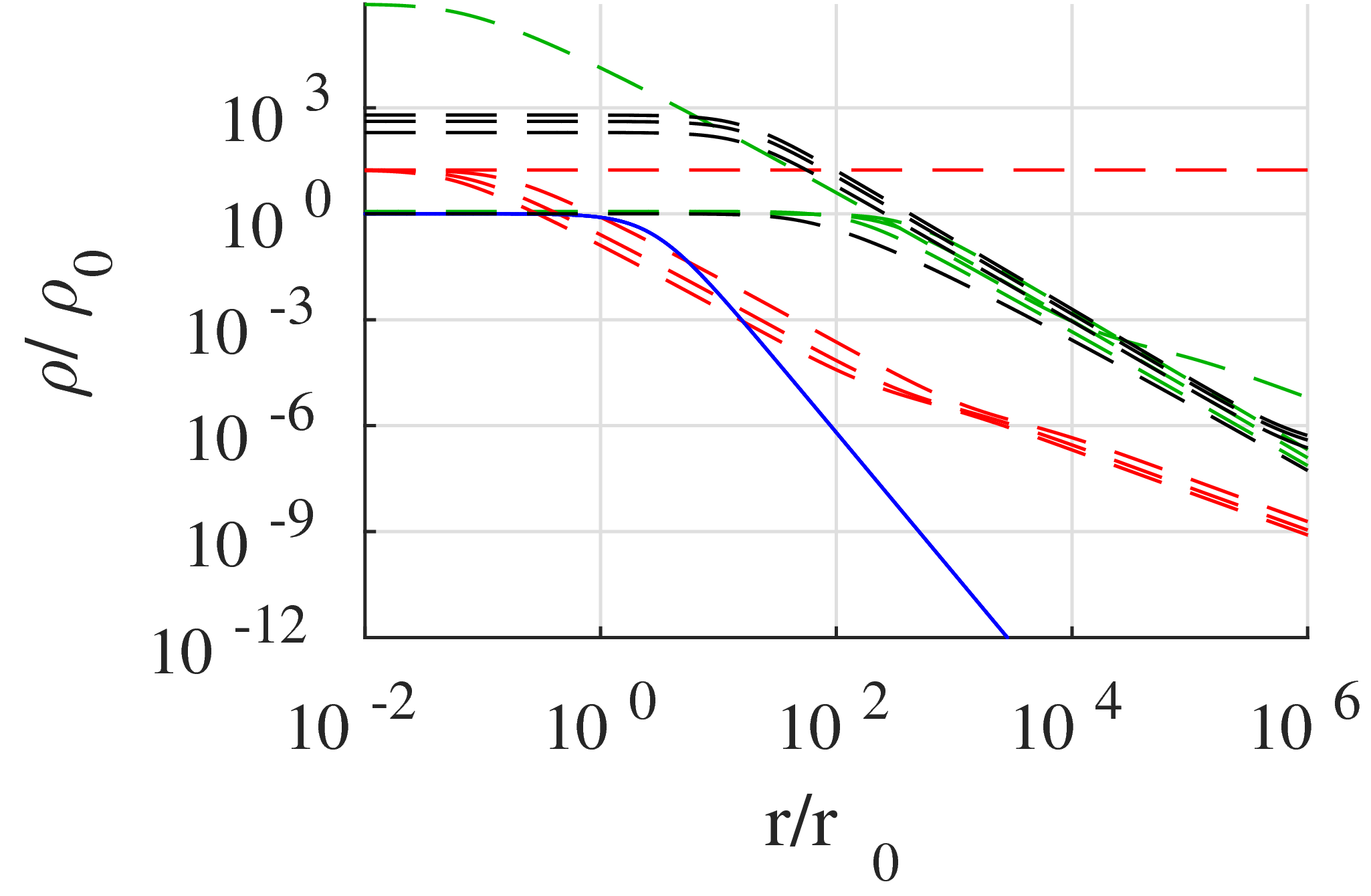}{0.4\textwidth}{}
          }                    
\caption{\textbf{Sample of changing model parameters excluding self-gravity and the torsional Alfv\'{e}n wave condition}. All plots have parameters set such that the dashed line colours corresponding to parameters that change linearly (step size of 4) in the range $\Omega_{0}=[0,3]$, $\Gamma_{z}=[10^{-3},25]$, and $\Gamma_{\phi}=[0,25]$ . Green lines correspond to changes in $\Gamma_{z}$, red lines to $\Gamma_{\phi}$, and black lines to $\Omega_{0}$. If a parameter is constant then they are set to: $\Gamma_{z} = 1$, $\Gamma_{\phi} = \pi/50$, and $\Omega_{0}=\pi/25$. For a set of coloured profiles, as $\Gamma_{z}$ is increasing, the density profiles extend radially. As $\Gamma_{\phi}$ is increasing, the density profiles pinch (decrease in density overall). As $\Omega_{0}$ increases, the overall density increases. \textbf{(Top Row)} Rotation law follows the radial power law (equation \ref{eq:OmegaPowerLaw}). \textbf{(Top Left)} $\alpha = -0.25$. \textbf{(Top Right)} $\alpha = -0.75$. \textbf{(Bottom Row)} Rotation law follows the flux power law (equation \ref{eq:OmegaPhiB}). \textbf{(Bottom Left)} $\alpha_{M} = -0.25$. \textbf{(Bottom Right)} $\alpha_{M} = -0.75$. It should be noted that each plot is magnified for clarity and the integration is indeed from $r_{min}\leq r \leq r_{max}$. The boundary condition $\rho(r_{min})/\rho_{0}=1$ is always met. In the case of $\alpha_{M} = -0.75$, $\rho$ increases rapidly at $r/r_{0}<<1$ due to unrealistic, extreme rotation near the origin, which is outside of the plot frame. All profiles which exceed $\rho_{0}$ are rejected from analysis and are considered unphysical models. These density profiles are not truncated for the purpose of illustration only. \label{fig:DensityProfileCompare}}
\end{figure*}

Theoretical density profiles from solving equation  \ref{eq:11Oct2015(6)} (and \ref{eq:11Oct2015FA}) are shown in Figures \ref{fig:DensityProfilesTheoreticalPL} and \ref{fig:DensityProfilesTheoreticalPhiB}. We halt the integration if a density inversion is encountered ($d\rho/dr>0$), as recommended by \citet{Recchi2014}. This condition also limits the density so that $\rho(r)\leq\rho_{0}$ everywhere. Some density profiles may be asymptotically constant at large radii, so these profiles were noted accordingly, but were allowed to integrate over the full range of $r$. 
For models including self-gravity, constant angular frequency, constant flux-to-mass ratios, and excluding the torsional Alfv\'{e}n wave condition, changing $\Gamma_{z}$ changes the asymptotic density value that is reached at large radii only by a small amount. This asymptotic value of $\rho$ at large radii can be approximated from equation \ref{eq:11Oct2015(6)}, which reveals that
\begin{equation}
\rho \approx 2\pi \bigg(\frac{\Omega_{0}}{\Gamma_{\phi}}\bigg)^{2}.
\end{equation}
\noindent 
\par

Truncation pressures were chosen randomly, with half chosen from a uniform distribution, and half chosen from a logarithmic distribution. This choice allows us to sample a wide range of possible external pressures that a filament may reside in. The logarithmic distribution allows the truncation values to span the entire range of $r$, which is particularly useful for probing the low density tail that most density profiles exhibit at large $r$. The uniformly distributed truncation values better sample where observationally constrained models are more likely to reside (discussed further below; see Figure \ref{fig:3DScatterFAPL} as well). Solutions that exhibit density inversions are halted at the first density inversion, then further truncation pressures are chosen along the truncated density profile with the aforementioned distribution. At each point, a number of important parameters ($\rho$, $P_{S}$, $\Omega_{S}$, $\Gamma_{\phi}$, $\Gamma_{z}$, etc.) are evaluated. In total, we sample approximately $10^{4}$ random combinations of the model parameters (equation \ref{eq:PLParRange}), each describing a unique density profile for a particular rotation law. Additionally, each external truncation pressure characterizes a unique filament model. Since there are {approximately $10^{4}$} external pressures, we explore approximately $10^{8}$ models for a given rotation law.
\par

The parameters evaluated at each truncation pressure ($\rho$, $P_{S}$, $\Omega_{S}$, $\Gamma_{\phi}$, $\Gamma_{z}$, etc.) are used to constrain the parameter space where observed molecular filaments reside, according to the constraints of equation \ref{eq:NonDimPars}. By converting the observed parameters into dimensionless quantities via equation \ref{eq:NonDimPars} (shown explicitly in Section \ref{sec:PigtailDHNGCT}), we plot where real filaments reside in the dimensionless parameter space (see Figures \ref{fig:3DScatterFAPL} and \ref{fig:Figure4}). 
The density profiles are also categorized based on their general behaviour or special characteristics. For example, some theoretical density profiles feature inversions (see Figure \ref{fig:161008_fa1sg1_pl_alphan025_dp_inversion}), some reach a constant density (at large radii), and some increase above $\rho_{0}$ (see Figures \ref{fig:DensityProfilesTheoreticalPhiB} and \ref{fig:DensityProfileCompare}).
\par

\begin{figure*}
\figurenum{7}
\centering
\plotone{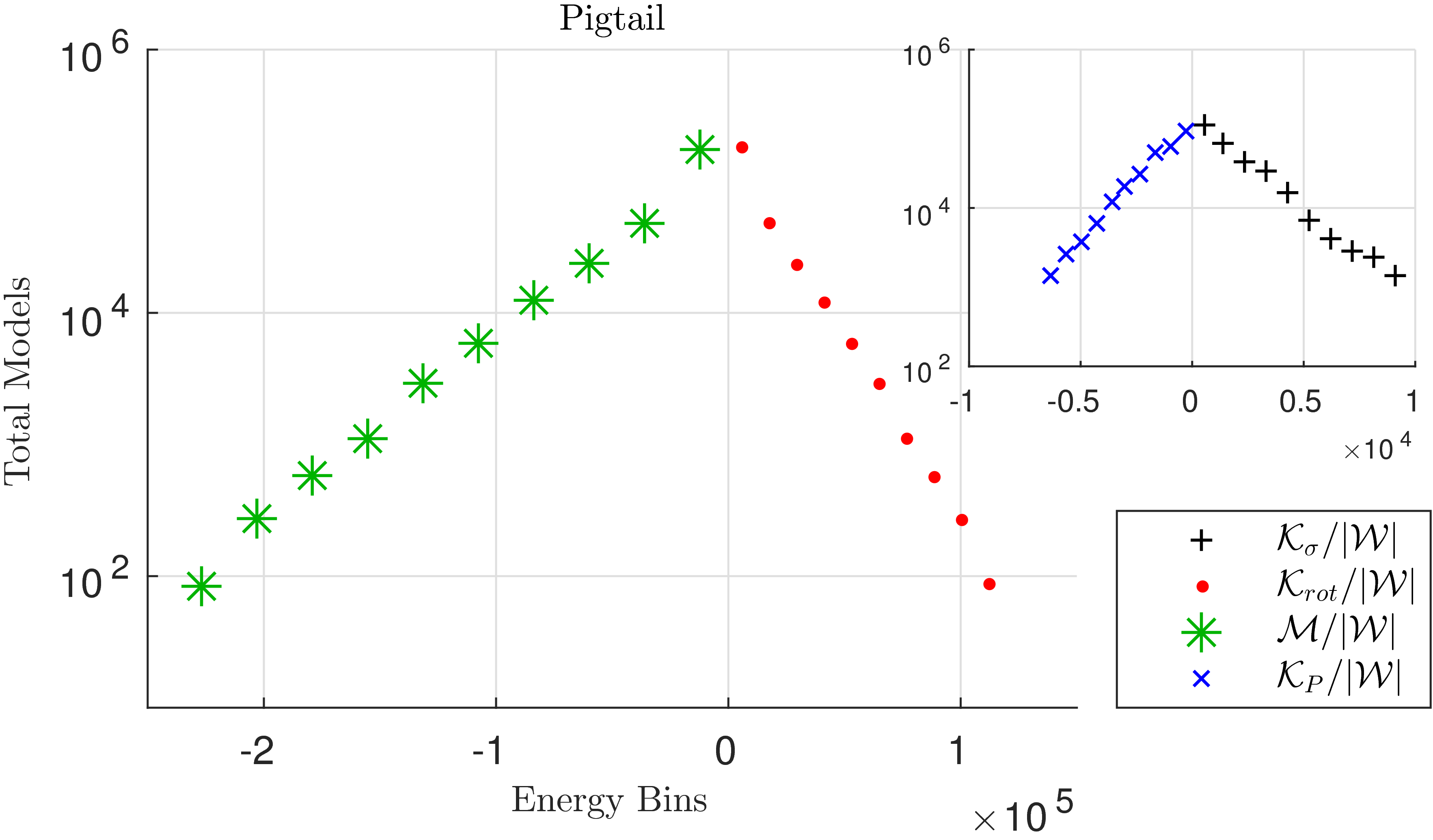}
\caption{\textbf{Relative energies of observationally constrained models including self-gravity, power law rotation ($\mathbf{\Omega=\Omega_{0}(r/r_{0})^{-0.25}}$), and the torsional Alfv\'{e}n wave condition.} Observational constraints are based on measurements of the Pigtail. The energy terms  are taken from equation \ref{eq:scriptMKPn1}. Notice that the pressure and magnetic stresses, in particular, play an important role in the equilibrium filament structure. The inset has the same axis labels as the host plot.   \label{fig:161019_fa1sg1_pl_alphan025_virial3D_linW_v3}}
\end{figure*}

\begin{figure}
\figurenum{8}
\centering
\plotone{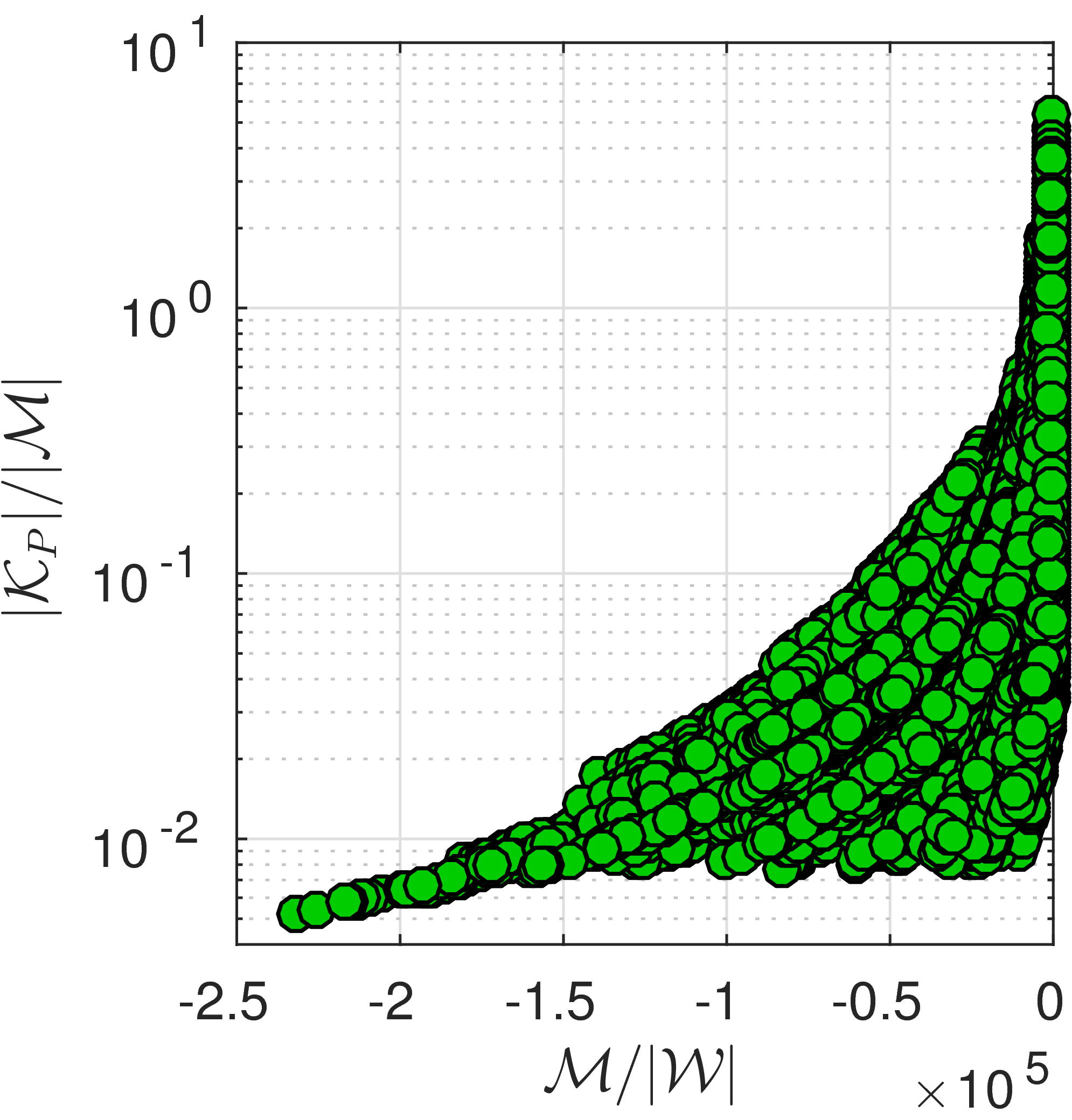}
\caption{A semi-logarithmic plot of $|\mathcal{K}_{P}|/|\mathcal{M}|$ against $\mathcal{M}/|\mathcal{W}|$ shows that most self-gravitating models that include the torsional Alfv\'{e}n wave condition (equation \ref{eq:FieldAngle2}) within observational constraints exhibit $|\mathcal{M}|>|\mathcal{K}_{P}|$. Quantities here are the same as those found in Figure \ref{fig:161019_fa1sg1_pl_alphan025_virial3D_linW_v3}.   \label{fig:161018_fa1sg1_pl_alphan025_virialOn_bsWOn_KPgeqMOn}}
\end{figure}

Having constrained the models within observational bounds, they can then be used to analyze the virial equation (equation \ref{eq:scriptMKPn1}). The result of this is shown in Figure \ref{fig:161019_fa1sg1_pl_alphan025_virial3D_linW_v3} for self-gravitating models following the radial power law (equation \ref{eq:OmegaPowerLaw}) with the torsional Alfv\'{e}n wave condition (equation \ref{eq:FieldAngle2}), constrained by the Pigtail. Within observational constraints, our analysis shows that the magnetic-to-gravitational energy ratio is 
\begin{equation}
\label{eq:VirialMWCompare}
-10^{5}\lesssim\frac{\mathcal{M}}{|\mathcal{W}|}\lesssim 0,
\end{equation}
and the ratio of the surface pressure term to gravitational energy is
\begin{equation}
\label{eq:VirialKPWCompare}
10^{0} \lesssim \frac{|\mathcal{K}_{P}|}{|\mathcal{W}|} \lesssim 10^{3}.
\end{equation}
Comparing the surface pressure term to the magnetic energy reveals that 
\begin{equation}\nonumber
|\mathcal{M}| > |\mathcal{K}_{P}|
\end{equation}
for most self-gravitating models with equation \ref{eq:FieldAngle2} within observational constraints of the Pigtail as shown in Figure \ref{fig:161018_fa1sg1_pl_alphan025_virialOn_bsWOn_KPgeqMOn}.
These results are discussed further in Section \ref{sec:IsSelfGravityImportant}.
\par

\section{Discussion}
\label{sec:Discussion}
In this section, we discuss the results presented in Section \ref{sec:Results}. We compare the behaviour of our models when varying different parameters, to the Ostriker solution. Models that exclude torsional Alfv\'{e}n wave physics (equation \ref{eq:FieldAngle2}) in Section \ref{sec:Excluding the Helical Field Angle Constraint} are first discussed, and in Section \ref{sec:Including the Helical Field Angle Constraint} we discuss models that include the constraints of torsional Alfv\'{e}n waves. The results of our virial analysis are discussed in Section \ref{sec:IsSelfGravityImportant}.

\subsection{Excluding the Torsional Alfv\'{e}n Wave Condition}
\label{sec:Excluding the Helical Field Angle Constraint}
Not surprisingly, the effect of self-gravity is to keep the filament together against the centrifugal force. In the absence of significant self-gravity, $\rho$ often tends to increase with radius and may exceed $\rho_{0}$ due to rotation, which is not realistic for a filament that eventually merges with the ISM. Several solutions of this type can be seen in Figures \ref{fig:DensityProfilesTheoreticalPhiB} and \ref{fig:DensityProfileCompare}. Typically, this behaviour is associated with strong rotation and weak toroidal magnetic fields. However, if magnetic stresses are strong enough, and/or rotation slows within a reasonable radius, then the density profile may decrease. Empirically, if a power law in radius (equation \ref{eq:OmegaPowerLaw}) is followed (see Figure \ref{fig:DensityProfileCompare}), then a greater $\Gamma_{\phi}/\Omega_{0}$ ratio tends to keep $\rho<\rho_{0}$ and $d\rho/dr<0$. The effect of $\Gamma_{z}$ does not seem to be as significant compared to $\Gamma_{\phi}$ and $\Omega_{0}$, unless $\Gamma_{z}$ is large. These features can be seen in Figure \ref{fig:DensityProfileCompare}. \par

If the flux power law of rotation (equation \ref{eq:OmegaPhiB}) is followed (see Figure \ref{fig:DensityProfileCompare}) with self-gravity neglected, then the value of $\alpha_{M}$ seems to have a dominating effect on the behaviour of $\rho(r)$. There are density profiles where $\rho>\rho_{0}$, regardless of the value of $\alpha_{M}$. A larger portion of these models exceed $\rho_{0}$ as $\alpha_{M}$ becomes steeper. All of these models, except where $\Gamma_{\phi} = 0$, decrease in density as $r$ becomes large. The $\Gamma_{\phi}/\Omega_{0}$ ratio has a similar effect as in the models that follow equation \ref{eq:OmegaPowerLaw}. See Figure \ref{fig:DensityProfileCompare} for the associated density profiles. \par

\subsubsection{Density Inversions}
\label{sec:DensityInversions}

Density inversions are found in some models where $\rho(r)<\rho_{0}$ (for all $r$), but curiously only if self-gravity is included, and the field angle constraint in equation \ref{eq:FieldAngle2} is excluded. An important difference between our models, and the models of \citet{Hansen1976} and \citet{Recchi2014} is the inclusion of magnetic fields. Even without the constraint of torsional Alfv\'{e}n waves (equation \ref{eq:FieldAngle2}), if there are constant poloidal and toroidal magnetic fields, then density inversions can still be observed as long as $\Gamma_{\phi}$ is weak. \par
It is easier to see the effects due to $\Omega_{0} = \Omega(r)$, $\Gamma_{\phi}$, and $\Gamma_{z}$, by first assuming that they are constant. The effect of increasing $\Gamma_{\phi}$ is to dampen the density inversions (see Figure \ref{fig:161008_fa1sg1_pl_alphan025_dp_inversion}). As $\Omega_{0}$ increases, the entire density profile along with the inversions, shifts to greater densities, and the inversion amplitude decreases slightly. The effect of increasing $\Gamma_{z}$ is that the density inversions tend to occur at larger radii as if each inversion was shifted over, and with larger amplitude. The toroidal flux-to-mass ratio $\Gamma_{\phi}$ seems to have a more noticeable effect on the behaviour of $\rho$ than $\Gamma_{z}$. Generally, $\Gamma_{\phi}$ and $\Gamma_{z}$ behave as \citet{Fiege1999a} describe -- $\Gamma_{z}$ tends to support/radially extend the filament compared to the Ostriker solution, while $\Gamma_{\phi}$ pinches the filament. These effects can still be seen even with density inversions occurring (see Figure \ref{fig:161008_fa1sg1_pl_alphan025_dp_inversion}).
\par
Density inversions exist in solutions where the rotation follows the radial power law of equation \ref{eq:OmegaPowerLaw} (see Figure \ref{fig:161008_fa1sg1_pl_alphan025_dp_inversion}), and for those that follow the flux power law of equation \ref{eq:OmegaPhiB}. At large radii, the density decreases like a power law for $-1 \leq \alpha < 0$. The general behaviour of the density inversions for both rotation laws is similar. There \textit{is} a subtle difference in that the inversions following the flux power law seem to exhibit a smaller frequency and features are stretched over a slightly larger radius than those that follow equation \ref{eq:OmegaPowerLaw}. While a particular model may exhibit density inversions, it is only realistic if it is pressure truncated before any inversions are observed.
\par

\subsection{Including the Torsional Alfv\'{e}n Wave Condition}
\label{sec:Including the Helical Field Angle Constraint}
Self-gravitating filaments following the radial power law of equation \ref{eq:OmegaPowerLaw} all have density profiles that decrease with radius (see Figure \ref{fig:DensityProfilesTheoreticalPL}).  A steeper power law index  tends to cause a steeper decline in density, as seen in Figure \ref{fig:DensityProfilesTheoreticalPL}. A large $\Gamma_{z}/\Omega_{0}$ ratio tends to produce steeper solutions initially, but then achieve an asymptotically constant ($d\rho/dr\approx 0$) value of $\rho$ at large radii. As $\alpha$ decreases, asymptotic solutions can be seen with smaller $\Gamma_{z}/\Omega_{0}$ ratios.
Non-self-gravitating filaments following the same rotation law generally have solutions that are not asymptotic. As $\alpha$ becomes progressively steeper, the density profiles tend to become more shallow. At $\alpha = -1$, all profiles are truly asymptotically constant at $\rho = \rho_{0}$. All profiles decrease when $\alpha > -1$ .
\par

Models that are self-gravitating and rotate according to the flux power law (equation \ref{eq:OmegaPhiB}) have decreasing density when $\alpha_{M} = -0.25$, but more negative values  have profiles that were deemed unrealistic because $\rho>\rho_{0}$ at some point(s) (see Figure \ref{fig:DensityProfilesTheoreticalPhiB}). All profiles begin to decrease at a sufficiently large radius, but the density inversion, where $d\rho/dr > 0$, becomes more pronounced as $\alpha_{M}$ becomes more negative. The existence of this extreme density inversion persists even after numerous tests of the numerical integration. Thus, we deem these models to be physically unrealistic and exclude them. As $\alpha_{M}$ becomes more negative, the profiles following the flux power law for rotation (equation \ref{eq:OmegaPhiB}) have a greater frequency of asymptotic solutions where there is a relatively large $\Gamma_{z}/\Omega_{0}$ ratio. The ratio of $\Gamma_{z}/\Omega_{0}$ does not need to be as large to observe asymptotic solutions as $\alpha_{M}$ decreases.
Non-self-gravitating filaments following the same rotation law do not have asymptotic solutions at $\alpha_{M} = -0.25$, and the density profiles decrease with the radius. Density profiles with $\alpha_{M} \lesssim -0.50$ were deemed unrealistic due to density inversions that increase the density above $\rho_{0}$ at some point(s).  This $\alpha_{M} \approx -0.50$ boundary coincides with the limit of $\alpha_{M}$ that was stated in Section \ref{sec:PhiB}. More negative values of $\alpha_{M}$ cause $\Omega$ to behave as $r^{-1}$ or steeper, and the filament rotates extremely rapidly near the origin. These models are deemed unrealistic.
\par

Generally, by including the physics of Alfv\'{e}n waves through the description in Section \ref{sec:FieldAngle}, we see that the density profiles are better behaved in the sense that the density profiles usually do not increase above $\rho_{0}$, the density falls off with the radius, and there are no density inversions. Interestingly, every model that follows the radial power law rotation of equation \ref{eq:OmegaPowerLaw} exhibit $\rho(r)\leq\rho_{0}$ (or are at least asymptotically constant at $\rho = \rho_{0}$ when $\alpha = -1$ for non-self-gravitating models), in both self-gravitating and non-self-gravitating cases regardless of any $\alpha < 0$. Furthermore, by including the Alfv\'{e}n wave constraint (equation \ref{eq:FieldAngle2}), every model that follows the flux power law rotation of equation \ref{eq:OmegaPhiB} also exhibit $\rho(r)\leq\rho_{0}$ in both self-gravitating and non-self-gravitating cases, as long as $\alpha_{M}>-1/2$. This suggests that the inclusion of Alfv\'{e}n wave physics is an important factor in the description of molecular tornadoes.
\par

\subsection{Is Self-Gravity Important?}
\label{sec:IsSelfGravityImportant}
Self-gravity, magnetic fields, and external pressure have a significant influence on constricting the density profile in our models. On the other hand, rotation, and turbulence tend to increase the density with radius. Interestingly, the observationally constrained models (see Figure \ref{fig:161018_fa1sg1_pl_alphan025_virialOn_bsWOn_KPgeqMOn}) indicate via equation \ref{eq:scriptMKPn1} that, for  all models,  $\mathcal{M}/|\mathcal{W}| < 0$. This is not surprising since we have noticed the pinching effect that magnetic fields have on density profiles (see Section \ref{sec:DensityInversions}, Figure \ref{fig:DensityProfileCompare}), and the magnetic-squeezing/twisting mechanism that is proposed by \citet{Matsumura2012,Morris2006,Sofue2007}.
\par

Having found the relative energies via our virial analysis (see Section \ref{sec:Results}), $\mathcal{W}$ is compared to $\mathcal{M}$ and $\mathcal{K}_{P}$, which are the quantities that can help keep the filament structure bound, to see which terms are particularly important. An example of these comparisons is shown in Figure \ref{fig:161019_fa1sg1_pl_alphan025_virial3D_linW_v3} for the self-gravitating model with the torsional Alfv\'{e}n wave constraint of equation \ref{eq:FieldAngle2}, and radial power law of equation \ref{eq:OmegaPowerLaw} with $\alpha = -0.25$. The magnetic energy ratio with gravitational energy within observational constraints is found to be $-10^{5}\lesssim \mathcal{M}/|\mathcal{W}|\lesssim 0$, which suggests that the magnetic field may be much more important than self-gravity. Furthermore, the ratio of the surface pressure to gravitational energy across all self-gravitating models within observational constraints is $10^{0} \lesssim |\mathcal{K}_{P}|/|\mathcal{W}| \lesssim 10^{3}$, which further indicates that these molecular tornadoes are only weakly self-gravitating. It is also found that $|\mathcal{M}|>|\mathcal{K}_{P}|$ for most models within observational constraints, as shown in Figure \ref{fig:161018_fa1sg1_pl_alphan025_virialOn_bsWOn_KPgeqMOn}. Thus, the toroidal magnetic field confinement is more important than the surface pressure, and both are more important than self-gravity. This result is similar to the interpretation by \citet{Sofue2007}, who considered the magnetic tension to be a significant component, and did not focus on the external pressure. Knowing this, the approximation $\mathcal{W}\approx 0$ may be justified. If the toroidal magnetic stress is dominant, equation \ref{eq:scriptMKPn1} may best be rewritten as 
\begin{equation}\label{eq:VirialEquationNonSG2}
\frac{|\mathcal{K}_{P}|}{\mathcal{M}}=\frac{\mathcal{K}_{rot}+\mathcal{K}_{\sigma}}{\mathcal{M}}+\frac{1}{2},
\end{equation}
\noindent where it is easier to see that if $\mathcal{M}<0$ then $(\mathcal{K}_{rot}+\mathcal{K}_{\sigma})<-\mathcal{M}/2$, and if $\mathcal{M}>0$ then $(\mathcal{K}_{rot}+\mathcal{K}_{\sigma})>-\mathcal{M}/2$. It was also found that $|\mathcal{M}|>|\mathcal{K}_{P}|$, and since  models follow $\mathcal{M}<0$, this means that $-(1/2)|\mathcal{M}|<(\mathcal{K}_{rot}+\mathcal{K}_{\sigma})<-(3/2)|\mathcal{M}|$.

\begin{figure*}
\figurenum{9}
\gridline{\includegraphics[width=0.5\textwidth]{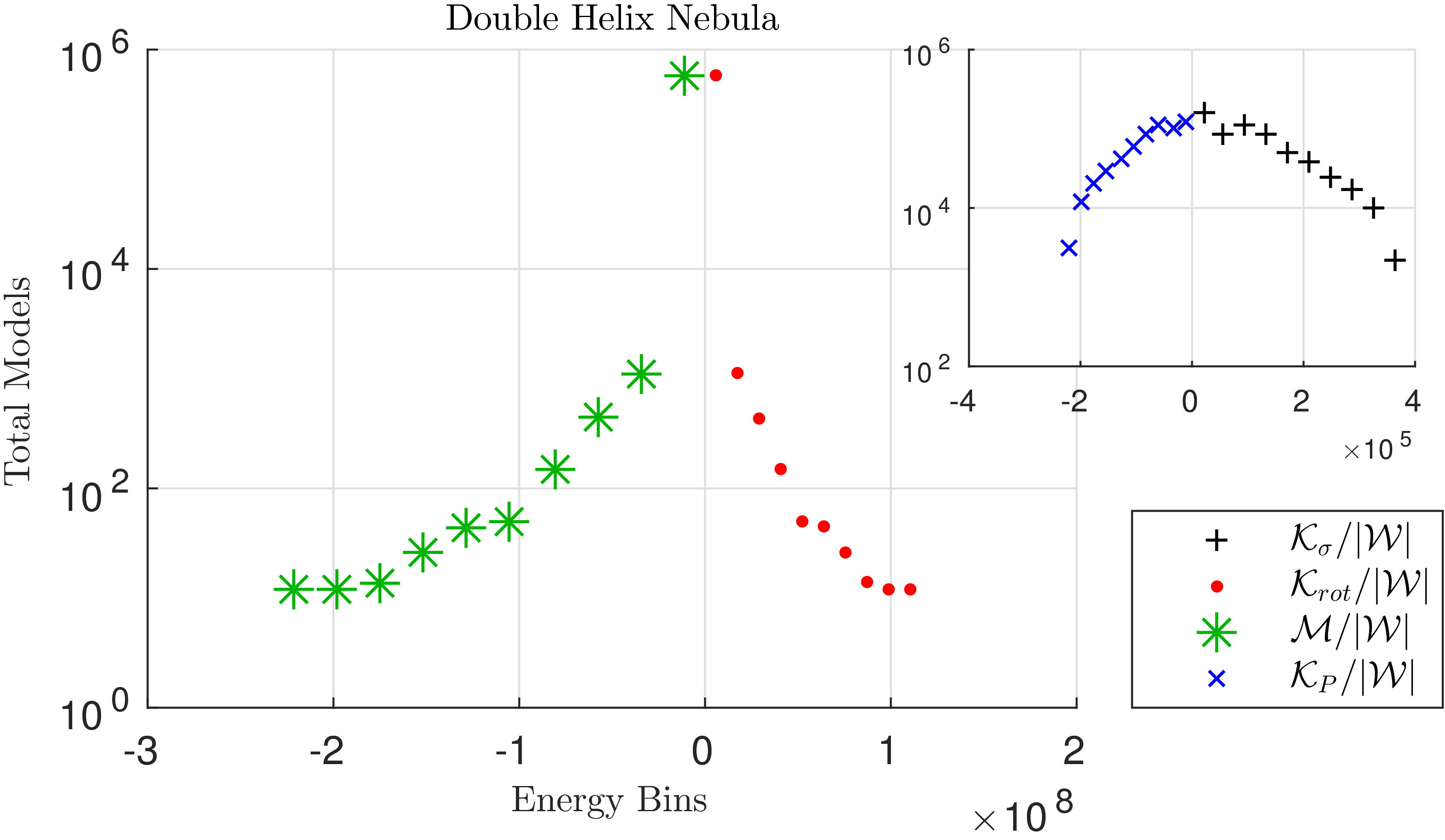}
          \includegraphics[width=0.5\textwidth]{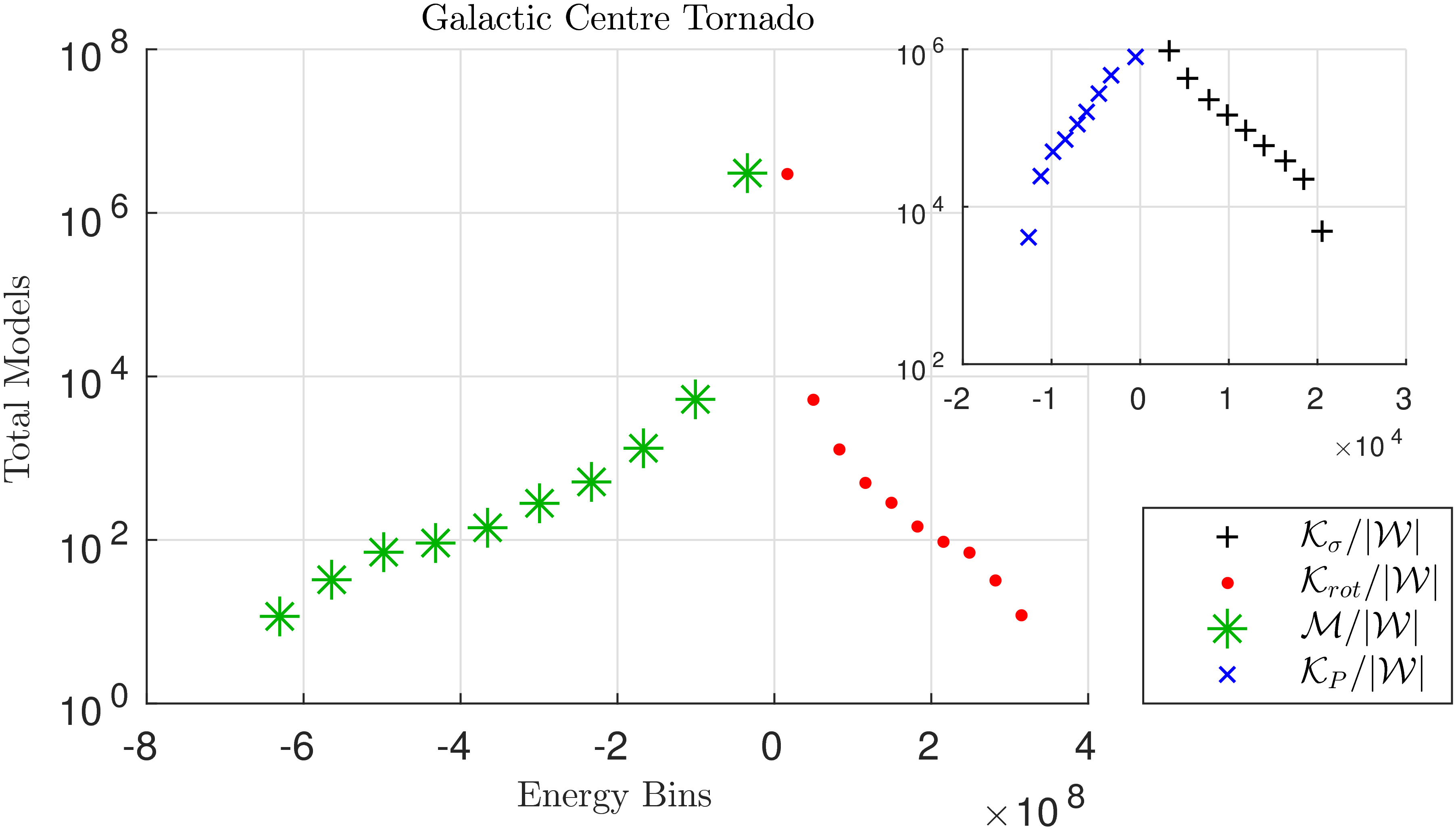}
          }
\caption{DHN (left) and GCT (right) histograms analogous to Figure \ref{fig:3DScatterFAPL}.  \label{fig:DHNGCT}}
\end{figure*}

A similar analysis of the DHN and GCT is shown in Figure \ref{fig:DHNGCT}.  All observationally constrained models of the GCT are dominated by the toroidal magnetic stress component, and the vast majority of models are magnetically dominated over the pressure. This lends support for the magnetic squeezing model by \citet{Sofue2007}. Analysis suggests that the dominant component of the DHN is not clear -- pressure or magnetic stresses may dominate. Comparison between $|\mathcal{M}|$ and $|\mathcal{K}_{P}|$ for the DHN reveals that more of the constrained models are magnetically dominated, though the dominating component cannot be conclusively determined. This uncertainty is likely due to estimating $\Omega_{S}$ By the torsional Alfv\'{e}n wave condition (equation \ref{eq:FieldAngle2}), and as seen in Figures \ref{fig:161019_fa1sg1_pl_alphan025_virial3D_linW_v3} and \ref{fig:DHNGCT}, rotation and the toroidal magnetic stress are directly proportional. A more accurate method of estimating the equilibrium filament radius would help constrain $\Omega_{S}$, and therefore, determine whether magnetic or pressure terms dominate. If the rotation had instead been estimated by the Shafranov condition (equation \ref{eq:Shafranov1958_stability_m=2}), then a much faster rotation would have been resulted. This would increase $|\mathcal{K}_{rot}|$, and would then require a much stronger $|\mathcal{M}|$ to satisfy the virial relation. Thus, the DHN would convincingly be magnetically dominated. It is highly suggestive, however, that both the DHN and GCT are weakly self-gravitating, like the Pigtail Molecular Cloud. We do not analyze the results of the DHN and GCT in detail as the observational constraints carry significant uncertainty associated with observational measurements.
\par

\section{Summary}
\label{sec:Summary}
Molecular tornadoes are fascinating objects that reside in the extreme environment of the CMZ. We study equilibrium MHD models of molecular tornadoes and introduce different rotation laws (see Section \ref{sec:DifferentialRotation}), reasonably general helical magnetic fields (equations \ref{eq:FluxToMass} and \ref{eq:11Oct2015(6)}), external pressures, and consider an isothermal equation of state.
\par

\begin{enumerate}

\item The physics of torsional Alfv\'{e}n waves was introduced in Section \ref{sec:FieldAngle}, which led to a constraint on the magnetic field components (equation \ref{eq:FieldAngle1}), and between $\Gamma_{\phi}$ and $\Omega$ (equation \ref{eq:FieldAngle2}). This constraint leads to solutions that are generally more well-behaved than solutions excluding the torsional Alfv\'{e}n wave condition -- in the sense that their density profiles do not increase indefinitely, and there are no density inversions. We conclude that including torsional Alfv\'{e}n wave physics is an important component in a realistic model of molecular tornado structure.

\item A special analytical solution, where centrifugal forces balanced exactly with toroidal magnetic stresses, was explored. The solution also assumes that the ratio between the pressure (from the isothermal equation of state) and the magnetic stresses, is constant (equation \ref{eq:constantbeta1}, and \ref{eq:constantbetacomponents}). These assumptions leads to solutions that are rescaled versions of the Ostriker solution, with constant rotational velocity of the filament (equation \ref{eq:11June2015Ostriker}). 

\item A Monte Carlo analysis of our models was conducted to explore the associated parameter space. Our models are constrained by limited observable constraints via equation \ref{eq:NonDimPars}. We focus our analysis on the Pigtail Molecular Cloud whose observational properties are more conclusively measured. Our analysis suggests that external pressures are important in its equilibrium structure (see Figure \ref{fig:161018_fa1sg1_pl_alphan025_virialOn_bsWOn_KPgeqMOn}). 

\item We performed a study of the virial theorem and found that, within observational constraints for the Pigtail,
\begin{flalign*}
&&-10^{5}\lesssim &\frac{\mathcal{M}}{|\mathcal{W}|}\lesssim 0,	&\\
&& 10^{0} \lesssim &\frac{|\mathcal{K}_{P}|}{|\mathcal{W}|} \lesssim 10^{3}, &\\
\text{and} && |\mathcal{M}| &> |\mathcal{K}_{P}|.
\end{flalign*}
Thus, the the confinement of magnetic fields dominates external pressure, which dominates self-gravity. We conclude that self-gravity is relatively unimportant in the equilibrium structure of molecular tornadoes. The relatively large values of $\mathcal{M/|W|}$ and $|\mathcal{K}_{P}|/|\mathcal{W}|$ are due to the weakness of self-gravity. Since $\mathcal{M}/|\mathcal{W}|< 0$, the magnetic field is dominated by the toroidal component, which lends support for the proposed magnetic-squeezing/twisting mechanism of molecular tornadoes.

\item We performed the same analysis for the DHN and GCT as we did for the Pigtail, but the conclusions we draw should be taken with more caution for these objects. Our analysis seems to suggest, however, that the GCT is dominated by the toroidal magnetic stress, whereas the DHN may be dominated by the pressure and/or magnetic stress. Self-gravity is relatively unimportant in these molecular tornadoes as well.

\end{enumerate}

\section{Acknowledgments}
K.A. would like to acknowledge Tomoharu Oka, Mark Morris, and Yoshiaki Sofue for their insights and interpretations on the Pigtail Molecular Cloud, DHN, and GCT, respectively. J.D.F. acknowledges the support of a Discovery Grant from the National Sciences and Engineering Research Council of Canada.

\appendix
\section{Torsional Alfv\'{e}n Wave Condition via Small Perturbations}\label{sec:SmallOscillationTorsionalAlfvenWave}
We assume a torsional Alfv\'{e}n wave of the form ${B}_{\phi} = B_{\phi,0}e^{-i(kz-\omega t)}$ and ${v}_{\phi} = v_{\phi,0}e^{-i(kz-\omega t)}$, where $B_{\phi,0}$ and $v_{\phi,0}$ are the corresponding amplitudes, $k$ is the wave number, and $\omega$ is the angular frequency of the perturbation. Notice that the differential operators $\partial_{t}$ and $\bm{\nabla}$ become $\partial_{t} \rightarrow i\omega$ and $\bm{\nabla} \rightarrow -ik\hat{z}$ when they act on the perturbed terms. By linearizing Faraday's Law, it becomes 

\begin{equation}\label{eq:AmperesLawtmp3}
\omega B_{\phi} = -kB_{z}v_{\phi}.
\end{equation}

\noindent Since the perturbations are small, it can be shown from the wave equation (equation \ref{eq:WaveEquation}) that 

\begin{equation}
v_{A}^{2} = \frac{\omega^{2}}{k^{2}},
\end{equation}
so equation \ref{eq:AmperesLawtmp3} can be simplified to

\begin{equation}
\label{eq:FieldAngle0}
\frac{B_{\phi,1}}{B_{z,0}} = \frac{v_{\phi,1}}{v_{A}},
\end{equation}

\noindent which is identical to equation \ref{eq:FieldAngle1}. Thus, the torsional Alfv\'{e}n wave condition (equation \ref{eq:FieldAngle2}) follows.

\bibliographystyle{aasjournal}
\bibliography{KA_JDF_1}

\listofchanges
\end{document}